\documentclass{ws-mpla}
\usepackage[super]{cite}
\usepackage{graphicx}
\begin{document}

\markboth{Authors' Names}
{Instructions for Typing Manuscripts (Paper's Title)}

\catchline{}{}{}{}{}

\title{Dynamical analysis of Brans-Dicke Universe with inverse power-law effective potential}

\author{\footnotesize Jonghyun Sim}

\address{Department of Physics, Soongsil University, Seoul 06978, Korea \\
ssujhsim@gmail.com}

\author{Jiwon Park}

\address{Department of Physics, Soongsil University, Seoul 06978, Korea}

\author{Tae Hoon Lee}

\address{Department of Physics, Soongsil University, Seoul 06978, Korea
             \and
             Research Institute for Origin of Matter and Evolution of Galaxies, Soongsil University, Seoul 06978, Korea
}

\def\bea{\begin{eqnarray}}
\def\eea{\end{eqnarray}}
\newcommand{\cl}{\centerline}
\newcommand{\na}{\nabla}
\def\beq{\begin{equation}}
\def\eeq{\end{equation}}
\def\pr{\prime}
\def\pa{\partial}
\def\un{\underline}
\def\ti{\tilde}
\def\na{\nabla}
\newbox\pippobox
\def\D{\Delta}
\newcommand{\nn}{\nonumber}
\def\o{\omega}
\def\M{\mu}
\def\N{\nu}
\def\vp{\varphi}
\def\pa{\partial}
\def\d{\delta}
\def\a{\alpha}
\def\b{\beta}
\def\c{\gamma}

\newbox\pippobo

\maketitle

\pub{Received (Day Month Year)}{Revised (Day Month Year)}

\begin{abstract}
We study Brans-Dicke cosmology with an inverse
power-law effective potential. By using dynamical analyses,
we search for fixed points corresponding to the radiation-like matter 
and dark energy-dominated era of our Universe, and the stability of  fixed points is also investigated. We find phase space trajectories which are attracted to the stable point of the dark energy-dominated era from  unstable fixed points like matter-dominated era of the Universe. The dark energy comes from effective potentials of the  Brans-Dicke field, whose variation (related to the time-variation of the gravitational coupling constant) is shown to be  in good agreement with observational data.

\end{abstract}

\ccode{PACS Nos.: include PACS Nos.}

\section{Introduction}

The recent acceleration of our Universe is thought  to be caused by the  mysterious dark energy, which composes  about 68\% \cite{1b, 2b, 3b, 4b}. Roughly 27\%  of the Universe consists of dark matter \cite{5b, 6b} and the remainder ordinary matter. One of the simplest candidate for dark energy is the well-known  cosmological constant.  The so-called  $\Lambda$CDM model is consistent with the current observational data \cite{7b}. Nonetheless, there still remain  fine tuning problems \cite{8b} like the cosmological constant  \cite{9b} and the (anthropic) cosmic coincidence problem \cite{10b} to be understood. 
 To suppress these problems, researchers have studied alternative models such as quintessence \cite{11b}, k-essence \cite{12b}, tachyon \cite{13b}, scalar-tensor theories including Brans-Dicke gravity \cite{14b}, and other theories. (See Ref.[15] and Ref.[16] for reviews of these models.) 

 In the standard model of particle physics Higgs-like fields have been studied to explain the primordial \cite{17b} and the late-time acceleration of the Universe \cite{18b}, and extended Higgs models containing the Einstein tensor  
 coupled, kinetic energy term have been examined \cite{19b, 20b, 21b}. Scalar-tensor theories have been also studied to explain the late-time acceleration of the Universe \cite{22b, 23b}.  
 Specifically, the recent acceleration of  the Universe could be explicated by the scalar field responsible for the early inflation, which  is a
 quintessence having an exponential potential or an inverse power-law potentials  \cite{24b, 25b, 26b, 27b, 28b, 29b, 30b, 31b, 32b}. They might be most  viable candidates to alleviate the coincidence problem. However, such potentials are not computed from a fundamental principle but are given by hand. 
  The dynamical analysis is an useful method to treat autonomous system, while comparing with the observational data about the dark energy and so on. The method  has been applied to scalar-tensor theories like Brans-Dicke gravity in Refs. [33 - 42] to describe the early or the late-time Universe. 
  (See Refs.[43 - 68] for other applications.) 

 In this paper, as in Ref. [69] we consider Brans-Dicke gravity with mutual interactions of the Brans-Dicke field  and a heavy  field. 
 In Sect. 2, we derive a low-energy effective potential \cite{50b, 51b} of the Brans-Dicke field, when the temperature of our Universe is much lower than the heavy field mass. 
In Sect. 3, we set up our model to analyze the Brans-Dicke Universe   as a dynamical system and find fixed points with various cosmological parameters. 
 %
In Sect. 4, with the inverse power-law effective potential we analyze  the dynamical system for cases  of some $\o$-values and investigate the stability around the fixed points. Also, with invariant submanifolds we reanalyze the dynamical system and investigate the stability around the fixed points. In Sect. 5, we study the de Sitter case of a specific fixed point to describe the late-time Universe.  In Sect. 6, we summarize our results.
 


   

\section{Effective potential}
In this section, we  briefly review the derivation of an effective potential from a high-energy theory by means of the low-energy effective theory formalism \cite{49b, 50b, 51b}. We consider the action for a high-energy theory
\begin{eqnarray}
S(\phi, h) = \int d^4 x \sqrt{-g}[\phi^2 R - \omega g^{\a \b} \pa_{\a} \phi \pa_{\b} \phi -V(\phi)]+S_\mathrm{m}, \nonumber  \\ 
\end{eqnarray}
\begin{eqnarray}
 S_{m}(\phi, h) = \int d^4 x \sqrt{-g}[-\frac{1}{2} g^{\a \b} \pa_{\a} h \pa_{\b} h -\overline{V}(h) + u\phi^k h^l] \\ \nonumber + \int d^4 x \sqrt{-g} L_\mathrm{om}.
\end{eqnarray}
Here $\o$ is related to the original Brans-Dicke coupling constant $\o_\mathrm{bd}$ as $\o = 4 \o_\mathrm{bd}$ \cite{14b}, $h$ is a heavy field, and $\phi$ is the Brans-Dicke field playing the role of a light field in the low-energy effective field theory. $L_\mathrm{om}$ is the lagrangian for the other matter. We consider the potential for a (Higgs-like) heavy field, $\overline{V}(h) = \frac{m_h^2 h^2}{2} + \frac{\lambda h^4}{4}$, and the second last term in Eq.(2) is an interaction between the heavy field  and the light Brans-Dicke field. 

When the freedoms
associated with a heavy feld are concealed from direct observation at a late-time of the  Universe of temperature  lower  than  the heavy field mass, within the tree-level approximation we have the following equation by applying the low-energy effective theory formalism \cite{49b, 50b, 51b} to Eqs. (1) and (2). 
\begin{eqnarray}
\frac{1}{\sqrt{-g}} \frac{\delta S(\phi, h)}{\delta h} = g^{\a \b} \nabla_{\a} \pa_{\b} h - \overline{V} ' (h) + \mathrm{u}l \phi^k h^{l-1} = 0. 
\end{eqnarray}
In the low-energy limit  $ \pa_{\b} h \ll m_h^2 h $, we can obtain $h(\phi)$ from Eq. (3) and an effective potential $ V_{eff}(\phi) = \frac{m_h^2 h^2(\phi)}{2} - \mathrm{u} \phi^k h^l (\phi) $ dependent on the Brans-Dicke field only (when $\lambda=0$). In the case   of the renormalizable interaction term \cite{49b},  with $k=1$ and $l=3$, $h(\phi)$ and $V_\mathrm{eff}(\phi)$ can be written as
\begin{eqnarray} 
&& h(\phi) \simeq \frac{m_h^2}{3 \mathrm{u} \phi} , \\ 
&& V_\mathrm{eff}(\phi) \simeq \frac{m_h^6 }{54 \mathrm{u}^2 \phi^{2}}.
\end{eqnarray}
Consequently, from Eqs. (1)-(5) we obtain the low-energy effective theory action depenent on the Brans-Dicke field  $\phi$ only and the other matter. 
\begin{eqnarray}
S(\phi,h(\phi)) = \int d^4 x \sqrt{-g}[\phi^2 R - \omega g^{\a \b} \pa_{\a} \phi \pa_{\b} \phi -V(\phi)] \ \ \ \\ \nonumber + \int d^4 x \sqrt{-g}[-\frac{1}{2} g^{\a \b} \pa_{\a} h(\phi) \pa_{\b} h(\phi) -\overline{V}(h(\phi))+u\phi h^3 (\phi)] \\ \nonumber + \int d^4 x \sqrt{-g} L_\mathrm{om}.
\end{eqnarray}

\section{Set up autonomous system}

In the flat Friedmann-Robertson-Walker (FRW) metric, $g_{\mu \nu}$ $=$ $Diag. (-1, a^2(\tau),$ $a^2(\tau), a^2(\tau))$ with a scale factor $a(\tau)$, the equations derived from Eq. (6) are given by
\begin{eqnarray}
&& 3H^2 = \frac{\rho}{2 \phi^2}, \\ && -(2 \dot{H} + 3 H^2) = \frac{p}{2 \phi^2}, \\ &&
2 \phi R + 2 \o (- \ddot{\phi} - 3 H \dot{\phi}) -V(\phi)_{, \phi} \\&& \nonumber - \ddot{h} h(\phi)_{, \phi} - 3 H \dot{h}(\phi) h(\phi)_{, \phi} - V_{eff}(\phi)_{, \phi} = 0,
\end{eqnarray}
where $H= \frac{\dot{a}(\tau)}{a(\tau)}$,  the dot, \  $\dot{}$ \, , denotes a derivative with respect to the cosmic time $\tau$, and $V(\phi)_{,\phi} \equiv \frac{d V(\phi)}{d \phi}$. The total energy density and pressure can be written as \cite{49b}
\begin{equation}
\rho =  \rho_\mathrm{bd}+ \rho_\mathrm{eff}+\rho_\mathrm{om} \ \ , \ \ p = p_\mathrm{bd}+ p_\mathrm{eff}+ p_\mathrm{om} ,
\end{equation}
where
\begin{eqnarray}
&& \rho_\mathrm{bd} = \o \dot{\phi}^2 -12 H \phi \dot{\phi} + V(\phi), \nonumber  \\ \nonumber && p_\mathrm{bd} = \o \dot{\phi}^2 + 4 (\dot{\phi}^2 + \phi \ddot{\phi} + 2 H \phi \dot{\phi}) -V(\phi),   \\ \nonumber 
&& \rho_\mathrm{eff} = \frac{1}{2} \dot{h}(\phi)^2 + V_\mathrm{eff}(\phi), \\&& p_\mathrm{eff} = \frac{1}{2} \dot{h}(\phi)^2 - V_\mathrm{eff} (\phi), 
\end{eqnarray}
 $\rho_\mathrm{om}$ is the energy density for the other matter,  and   $p_\mathrm{om}$ is the pressure.
Eq. (7)  can be rewritten as
\begin{eqnarray}
1= \frac{1}{6H^2 \phi^2} [\rho_\mathrm{om} + \o \dot{\phi}^2 - 12 H \phi \dot{\phi} + V +  \frac{1}{2} \dot{h}(\phi)^2 + V_\mathrm{eff}(\phi)]. \nonumber \\ 
\end{eqnarray}
With dimensionless variables 
\begin{eqnarray}
x^2 \equiv \frac{\dot{\phi}^2}{6 H^2 \phi^2}, \  y^2 \equiv \mathrm{c} \frac{\dot{\phi}^2}{H^2 \phi^6}, \ z^2 \equiv \frac{V(\phi)}{6 H^2 \phi^2}, \ t^2 \equiv \frac{V_\mathrm{eff}(\phi)}{6 H^2 \phi^2}, \nonumber \\ 
\end{eqnarray}
where the constant 
 $\mathrm{c}\equiv \frac{\mathrm{m_h}^4}{108 \mathrm{u}^2}$, Eq. (12) becomes
\begin{eqnarray}
1= \Omega_\mathrm{om} + \o x^2 - 2 \sqrt{6} x + z^2 +y^2 + t^2. \ \ \ \ \ \ 
\end{eqnarray}
 From Eqs. (8) and (9) we define other dimensionless variables 
  $A \equiv \frac{\ddot{\phi}}{H \dot{\phi}}$ and $B \equiv -\frac{\dot{H}}{H^2}$, which are dependent on each other as
\begin{eqnarray}
\frac{2}{3} B - w_\mathrm{om} \Omega_\mathrm{om} - (\o+4) x^2 - \frac{2}{3} \sqrt{6} A x \\ \nonumber - \frac{4 \sqrt{6}}{3} x + z^2 -y^2 +t^2 -1 =0,
\end{eqnarray}
\begin{eqnarray}
24 x -12 x B - 2 \sqrt{6} \o A x^2 - 6 \sqrt{6} \o x^2 -12 D z^2 x \\ \nonumber - 2 \sqrt{6} A y^2 +24 y^2 x - 6 \sqrt{6} y^2 -12 E t^2 x =0,
\end{eqnarray}
where $D = \frac{V'(\phi)}{2 V(\phi)} \phi $, $E = \frac{V_\mathrm{eff}'(\phi)}{2 V_\mathrm{eff}(\phi)} \phi$.

The ratio of the energy density of the other matter relative to $6H^2 \phi^2$ and that of the Brans-Dicke field can be expressed as 
\begin{equation}
\Omega_\mathrm{om}\equiv \frac{\rho_\mathrm{om}}{6H^2 \phi^2} = 1 - \o x^2 + 2 \sqrt{6} x - z^2  -y^2 - t^2 ,
\end{equation}
\begin{equation}
\Omega_{\phi} =  \o x^2 - 2 \sqrt{6} x + z^2 +y^2 + t^2 . 
\end{equation}
Eqs. (17) and (18) give us constraints, $0 \leq \Omega_{\phi} \leq 1$ and $0 \leq \Omega_\mathrm{om} \leq 1$. The  equation of state for the total energy and pressure and the equation of state regarding to the Brans-Dicke field are given by
\begin{eqnarray}
w_\mathrm{m}&& = \frac{p}{\rho}= \frac{p_\mathrm{om} + p_\mathrm{bd}+ p_\mathrm{eff}}{\rho_\mathrm{om} + \rho_\mathrm{bd}+ \rho_\mathrm{eff}} \ \ \ \ \ \\ \nonumber &&= w_\mathrm{om} \Omega_\mathrm{om}+(\o+4) x^2 + \frac{2 \sqrt{6}}{3}  x A + \frac{4 \sqrt{6}}{3} x - z^2 + y^2 - t^2 ,  \\  w_{\phi}&&=\frac{p_{\phi}}{\rho_{\phi}}= \frac{p_\mathrm{bd} + p_\mathrm{eff}}{\rho_\mathrm{bd} + \rho_\mathrm{eff}} \ \ \ \ \  \\ \nonumber  &&= \frac{(\o +4)x^2 + \frac{2 \sqrt{6}}{3}  x A  + \frac{4 \sqrt{6}}{3} x- z^2 + y^2 - t^2  }{\o x^2 - 2\sqrt{6} x + z^2 + y^2 + t^2} 
\end{eqnarray}
with $w_\mathrm{om} = \frac{p_\mathrm{om}}{\rho_\mathrm{om}}$.
Note that our Universe is accelerating if the  equation of state for the total energy and pressure $w_\mathrm{m} < - \frac{1}{3}$. If specially $w_\mathrm{m} = -1$, then the Universe must be accelerating because of the influence of the cosmological constant. On the other hand, if  $ - 1 < w_\mathrm{m} < - \frac{1}{3}$, then we have an accelerating Universe due to the presence of dark energy like quintessence. (With the equation of state for the total energy and pressure $w_\mathrm{m} =p/\rho$, 
 $\ddot{a}/a \, (=\dot{H}+H^2=-(\rho+3p)/(12\phi^2))
 =-(1+3\omega_\mathrm{m})\rho/(12\phi^2)$ and $\dot{H}=-(1+w_m)\rho/(4\phi^2)$ from Eqs. (7) and (8).)

Using Eqs. (12)-(16),
we can rewrite our autonomous system in Eqs. (7)-(9) as
\begin{eqnarray}
&& x' = x [A  + B - \sqrt{6} x],  \\ 
&& y' = y [A  + B - 3 \sqrt{6} x],  \\
&& z' = z [ \sqrt{6} D x + B - \sqrt{6} x], \\
&& t'  =  t [ \sqrt{6} E x + B - \sqrt{6} x], \\ 
&& D'= 2 \sqrt{6}  D^2 x [\frac{1}{2 D} + \Gamma -1], \\
&& E'= 2 \sqrt{6}  E^2 x [\frac{1}{2 E} +\Theta -1],
\end{eqnarray}
where $'$ denotes the derivative with respect to $N = $ln$a(\tau)$.
 We further define dimensionless variables as $\Gamma=\frac{V(\phi) V''(\phi)}{V'(\phi)^2}$, and $\Theta=\frac{V_\mathrm{eff}(\phi) V_\mathrm{eff}''(\phi)}{V_\mathrm{eff}'(\phi)^2}$. 
 
 In this paper, we take $V(\phi) \propto \phi^{\mathrm{n}}$ which is a power-law potential regarding to Brans-Dicke field and $V_\mathrm{eff}(\phi) = \frac{\mathrm{m_h}^6 M_\mathrm{P}^j}{54 \mathrm{u}^2 \phi^{2+j}}$ which is an inverse power-law potential derived by the low-energy effective theory formalism. 
 In this case, $D'=0$ and $E'=0$ since $D = \frac{\mathrm{n}}{2}$ and $E = -1 - \frac{j}{2}$ where $\mathrm{n}$ and $j$ are constant.\footnote{Note that if the scalar field with (inverse) power-law potentials is not the Brans-Dicke field, then we have to analyze 6D autonomous system because  $\frac{V'(\phi)}{\kappa V(\phi)}$ with $\kappa=\sqrt{8 \pi G}$
(and   $\frac{V_\mathrm{eff}'(\phi)}{\kappa V_\mathrm{eff}}$) that should be studied is dependent on $ \phi$ as in Ref. [28].
However in  Brans-Dicke gravity the Brans-Dicke field is related with the gravitational constant $G \propto  \frac{1}{\phi^2}$ \cite{14b}, and thus $D = \frac{V'(\phi)}{2 V(\phi)} \phi $ and $E = \frac{V_\mathrm{eff}'(\phi)}{2 V_\mathrm{eff}(\phi)} \phi$ are  constants in cases of (inverse) power-law potentials  \cite{34b, 35b, 42b,43b}. }

\subsection{Stability analysis of fixed points}

In this subsection, we determine the linear stability of a fixed point $(x=x_0, \ y=y_0, \ z=z_0, \ t=t_0)$ with a perturbation $x=x_0 + \delta x$, $y=y_0 + \delta y$, $z=z_0 + \delta z$, $t=t_0 + \delta t$ as 
\begin{displaymath}
\left( \begin{array} {c} \delta x' \\ \delta y' \\ \delta z' \\ \delta t'  \end{array}  \right) = M \left(  \begin{array}{c} \delta x \\ \delta y \\ \delta z \\ \delta t \end{array} \right),
\end{displaymath}
where $M$ is given by 
\begin{displaymath}
M	 =
\left(\begin{array}{cccc}
\frac{\partial{x'}}{\partial{x}} & \frac{\partial{x'}}{\partial{y}} & \frac{\partial{x'}}{\partial{z}} & \frac{\partial{x'}}{\partial{t}}  \\ \frac{\partial{y'}}{\partial{x}} & \frac{\partial{y'}}{\partial{y}} & \frac{\partial{y'}}{\partial{z}} & \frac{\partial{y'}}{\partial{t}}  \\ \frac{\partial{z'}}{\partial{x}} & \frac{\partial{z'}}{\partial{y}}  &  \frac{\partial{z'}}{\partial{z}} & \frac{\partial{z'}}{\partial{t}}  \\  \frac{\partial{t'}}{\partial{x}} & \frac{\partial{t'}}{\partial{y}} & \frac{\partial{t'}}{\partial{z}} & \frac{\partial{t'}}{\partial{t}}  \end{array} \right)_{(x=x_0, y=y_0, z=z_0, t=t_0)}
\end{displaymath}
to be calculated from Eqs. (21)-(24).
The above has four eigenvalues.
When all eigenvalues are negative, the fixed point is stable. When all eigenvalues are positive, the  fixed point is unstable. 
On the other hand, if some of eigenvalues are negative and the others positive, then the fixed point is saddle. If the determinant of the matrix $M$ is negative and  real parts of the eigenvalue are negative, then the fixed point is a stable spiral \cite{15b,16b,45b,47b}.  

 From Eqs. (15) and (16) we have
\begin{eqnarray}
A && = \frac{\sqrt{6}}{2(6 x^2 + \o x^2 +y^2)}\{  x  -3 w_\mathrm{om} x (1 - \o x^2 + 2 \sqrt{6} x - z^2 -y^2 -t^2)  \nonumber \\  && -3 \o x^3 - 12 x^3   - 4 \sqrt{6} x^2  + 3 x z^2 +3 x t^2 \\ \nonumber && - \sqrt{6} \o x^2 -2 D z^2 x + y^2 x - \sqrt{6} y^2 - 2 E t^2 x \}   ,  \\ 
B &&= \frac{3}{2} w_\mathrm{om} (1- \o x^2 + 2 \sqrt{6} x -z^2  - y^2 -t^2) \\ \nonumber&& + 6 x^2 + \sqrt{6} A x  + 2 \sqrt{6} x + \frac{3}{2} (-z^2 + y^2 -t^2 +1).
\end{eqnarray}

In sections $3$ and $4$, we analyze the $4$-dimensional dynamical system by investigating the fixed points and their stability and show some physically meaningful trajectories around fixed points. However, we find an useful mathematical method by which a dynamical system can be described more appropriately in low-dimensional phase spaces:  Invariant submanifolds are parts of entire phase space, which evolve to themselves under the dynamics, and each of them is not connected to any other areas  \cite{16b, key-1, key-3, key-6, key-10, key-15, key-17, key-19}. We can find invariant submanifolds by looking at the structure of our dynamical system of Eqs. (21) - (24) so that $x=0$ (without the kinetic term), $y=0$ (without the effective kinetic term), $z=0$ (without the potential term), or $t=0$ (without the effective potential term), also the vacuum case $\Omega_\mathrm{om} = 0$ is an invariant submanifold, respectively. This implies that a global attractor exists when $x=y=z=t=0$, but we cannot determine whether our dynamical system has a global attractor since a divergent singularity appears in Eqs. $(27)$ and $(28)$ (when $x=0$ and $y=0$). 
 Also, in section 5, assuming that $B=0$ and   $A \neq 0$ which  satisfy directly Eqs. (15) and (16) without using Eqs. (27) and (28), we investigate the stability of the fixed point, $(x_0=0, y_0=0)$, corresponding to the de Sitter Universe. 
  In subsections 3.2-3.5,  we summarize  various cosmological parameters of each fixed point obtained from Eqs. (21)-(24), the equation of state regarding to the Brans-Dicke field $w_{\phi}$, the total equation of state $w_{m}$, the density ratio of the Brans-Dicke field $\Omega_{\phi}$, and eigenvalues $\lambda^{i}$ with $i=1,2,3,4$.

\subsection{Fixed points  of $(x = 0, y \neq 0, z=0, t=0)$ type}
Among fixed points of Eqs. (21)-(24),  we have this type $(\mathrm{a})$.
\subsubsection{$(x_\mathrm{a}, y_\mathrm{a}, z_\mathrm{a}, t_\mathrm{a})$ =  $(0, \pm1, 0, 0) $}

This type of fixed points with $\Omega_{\phi_\mathrm{a}} =1$, $w_{\mathrm{m}_\mathrm{a}}=1$, and $w_{\phi_\mathrm{a}}=1$ describes  the stiff matter-dominated era of our Universe.  Eigenvalues are given by\\
$\lambda_\mathrm{a}^1 = 0, \lambda_\mathrm{a}^2 = 3-3 w_\mathrm{om}, \lambda_\mathrm{a}^3 = 3, \lambda_\mathrm{a}^4 = 3.$ \\
It is a normally unstable point (Non-Hyperbolic)\footnote{To complete analysis of stability for the fixed point where one of its eigenvalues is 0, we sholud consider a center manifold analysis \cite{key-1, key-6, key-10, key-14, key-15, key-16, key-17}. However, we don't analyze such a deeper dynamical analysis in present paper, which is denoted by Non-Hyperbolic.}.  

\subsection{Fixed points of $(x \neq 0, y=0, z=0, t=0)$ type}
Among fixed points of Eqs. (21)-(24), we obtain types $(\mathrm{b})$.

\subsubsection{ $(x_{\mathrm{b}1\pm},  y_{\mathrm{b}1},  z_{\mathrm{b}1},  t_{\mathrm{b}1}) = (\frac{\sqrt{6} \pm \sqrt{6+\o}}{\o},0,0,0)$}
 With $\Omega_{\phi_{\mathrm{b}1}} = 1$, $w_{\mathrm{m}_{\mathrm{b}1}} = \frac{24 + 3 \o \pm 4 \sqrt{6} \sqrt{6 + \o}}{3 \o}$, $w_{\phi_{\mathrm{b}1}} = \frac{24 + 3 \o \pm 4\sqrt{6} \sqrt{6 + \o}}{3 \o}$, eigenvalues are given by $ \lambda_{\mathrm{b}1 \pm}^1 = -\frac{2 (12 \sqrt{6} + \sqrt{6} \o \pm 12 \sqrt{6 + \o})}{\o (\sqrt{6} \pm \sqrt{6 + \o})},  \lambda_{\mathrm{b}1\pm}^2 = \frac{6 + 6 D + 3 \o \pm \sqrt{6} \sqrt{6 + \o} \pm  \sqrt{6} D \sqrt{6 + \o}}{\o}, \\  \lambda_{\mathrm{b}1 \pm}^3 = \frac{6 + 6 E + 3 \o \pm \sqrt{6} \sqrt{6 + \o} \pm  \sqrt{6} E \sqrt{6 + \o}}{\o},  \lambda_{\mathrm{b}1\pm}^4 = \frac{12 + 3 \o \pm 2 \sqrt{6} \sqrt{6 + \o} - 3 \o w_\mathrm{om}}{\o}. $ \\
For the  '$+$' case of fixed points $3.3.1$, stability conditions that all eigenvalues are negative are 
$E < -1$, $w_\mathrm{om} > \frac{4+\o}{\o}+\frac{2}{3}\sqrt{\frac{36 + 6 \o}{\o^2}}$ and  $D < 2 - \frac{1}{2}\sqrt{\frac{36 \o^2 + 6 \o^3}{\o^2}}$, $0 < \o < \frac{1}{18}(-30-24 E + 6 E^2)  + \frac{1}{18}\sqrt{900+1440 E + 216 E^2 - 288 E^3 + 36 E^4}$. The '$-$' case of the fixed points $3.3.1$ is unstable (saddle).

\subsubsection{$ (x_{\mathrm{b}2},  y_{\mathrm{b}2},  z_{\mathrm{b}2},  t_{\mathrm{b}2}) = (\frac{\sqrt{6} -3 \sqrt{6} w_\mathrm{om}}{12 + 3 \o -3 \o w_\mathrm{om}},0,0,0)$}
With $\Omega_{\phi_{\mathrm{b}2}} = -\frac{2 (-1 + 3 w_\mathrm{om}) (-24 + \o (-5 + 3 w_\mathrm{om}))}{3 (-4 + \o (-1 + w_\mathrm{om}))^2}$, $w_{\mathrm{m}_{\mathrm{b}2}} = \frac{\sqrt{6} -3 \sqrt{6} w_\mathrm{om}}{12 + 3 \o -3 \o w_\mathrm{om}}$, $w_{\phi_{\mathrm{b}2}} = \frac{(-8 + \o (-2 + w_\mathrm{om})) (-1 + 3 w_\mathrm{om})}{-24 + \o (-5 + 3 w_\mathrm{om})}$, eigenvalues are given by \\
$\lambda_{\mathrm{b}2}^1 = \frac{3 (-4 + \o (-1 + w_\mathrm{om})) (1 + w_\mathrm{om}) + 4 D (-1 + 3 w_\mathrm{om})}{-8 + 2 \o (-1 + w_\mathrm{om})}, \\  \lambda_{\mathrm{b}2}^2 = \frac{3 (-4 + \o (-1 + w_\mathrm{om})) (1 + w_\mathrm{om}) + 4 E (-1 + 3 w_\mathrm{om})}{-8 +  2 \o (-1 + w_\mathrm{om})}, \\  \lambda_{\mathrm{b}2}^3 = \frac{4 - 12 w_\mathrm{om}}{-4 + \o (-1 + w_\mathrm{om})},   
\lambda_{\mathrm{b}2}^4 = \frac{16 + 3 \o {(-1 + w_\mathrm{om})}^2 - 24 w_\mathrm{om}}{-8 + 2 \o (-1 + w_\mathrm{om})}. $ \\
 All the fixed points of 3.3.2 are unstable (saddle) with a constraint $0 \leq \Omega_{\phi_{\mathrm{b}2}} \leq 1$.

\subsection{Fixed points of $(x \neq 0, y=0, z \neq 0, t = 0)$ type}
Among fixed points of Eqs. (21)-(24), we have a type  $(\mathrm{c})$.

\subsubsection{$ (x_{\mathrm{c}1},  y_{\mathrm{c}1},  z_{\mathrm{c}1\pm},  t_{\mathrm{c}1}) = (-\frac{\sqrt{2/3} (-2 + D)}{(2 + 2 D + \o)}, 0, \pm \frac{ \sqrt{-(-10 - 8 D + 2 D^2 - 3 \o) (6 + \o)}}{
 \sqrt{3} (2 + 2 D + \o)}, 0)$}

With $\Omega_{\phi}=1$, $w_\mathrm{m}=\frac{2 - 18 D + 4 D^2 - 3 \o}{6 + 6 D + 3 \o}$, $w_{\phi} = \frac{2 - 18 D + 4 D^2 - 3 \o}{6 + 6 D + 3 \o}$, eigenvalues are given by
$\lambda_{\mathrm{c}1}^1 = -3 + \frac{ 2 (-2 + D) (1 + D)}{2 + 2 D + \o}, \lambda_{\mathrm{c}1}^2 = -3 + \frac{4 (-2 + D) D}{2 + 2 D + \o} - 3 w_\mathrm{om},  \\ \lambda_{\mathrm{c}1}^3 = \frac{2 (-2 + D) (D - E)}{2 + 2 D + \o}, \  \lambda_{\mathrm{c}1}^4 = \frac{4 (-2 + D)}{2 + 2 D + \o}.$ \\
For fixed points of 3.4.1, stability conditions that all eigenvalues are negative with $(-10 - 8 D + 2 D^2 - 3 \o)(6 + \o) < 0$ are $0 < D < 2$, $w_\mathrm{om} < -1$, $-2 -2 D < \o < \frac{-6 - 14 D + 4 D^2 - 6 w_\mathrm{om} - 6 D w_\mathrm{om}}{3 + 3 w_\mathrm{om}}$, $E < D$ or $0 < D < 2$, $w_\mathrm{om} \geq -1$, $\o > -2 -2 D$, $E < D$ or $D > 2$, $w_\mathrm{om} < -1$, $\frac{-6 - 14 D + 4 D^2 - 6 w_\mathrm{om} - 6 D w_\mathrm{om}}{3 + 3 w_\mathrm{om}} < \o < -2 -2 D$, $E < D$ or $D > 2$, $w_\mathrm{om} \geq -1$, $\o < -2 -2 D$, $E < D$.

\subsubsection{$ (x_{\mathrm{c}2},  y_{\mathrm{c}2},  z_{\mathrm{c}2\pm},  t_{\mathrm{c}2}) = (-\sqrt{\frac{3}{2}} \frac{(1 + w_\mathrm{om})}{2 D},0,  \pm \frac{\sqrt{F_{\mathrm{c}2}}}{2 \sqrt{2} D \sqrt{2 + 2 D + \o}},0)$}
\noindent
Here $F_{\mathrm{c}2} = D^2 (8 - 24 w_\mathrm{om}) - 
         3 (2 + \o) (-4 + \o (-1 + w_\mathrm{om})) (1 + w_\mathrm{om}) - 
         2 D (-16 + \o (-5 + 6 w_\mathrm{om} + 3 w_\mathrm{om}^2))$. With $w_{\mathrm{m}_{\mathrm{c}2}} = \frac{-1 + (-1 + D) w_\mathrm{om}}{D}$, $\Omega_{\phi_{\mathrm{c}2}}= \frac{3 (2 + \o) (1 + w_\mathrm{om}) + 2 D (7 + 3 w_\mathrm{om})}{4 D^2}$, \\
$w_{\phi_{\mathrm{c}2}} = \frac{3 (2 + \o) w_\mathrm{om} (1 + w_\mathrm{om}) + 2 D (-2 + 5 w_\mathrm{om} + 3 w_\mathrm{om}^2)}{3 (2 + \o) (1 + w_\mathrm{om}) + 2 D (7 + 3 w_\mathrm{om})}$, eigenvalues are given by\\
$\lambda_{\mathrm{c}2}^1 = \frac{3 D^2 (6 + \o) (2 + 2 D + \o) (1 + D (-1 + w_\mathrm{om}) + w_\mathrm{om}) - 
   \sqrt{3} \sqrt{F}}{4 D^3 (6 + \o) (2 + 2 D + \o)}, \\ 
\lambda_{\mathrm{c}2}^2 = \frac{3 D^2 (6 + \o) (2 + 2 D + \o) (1 + D (-1 + w_\mathrm{om}) + w_\mathrm{om}) + 
   \sqrt{3} \sqrt{F}}{4 D^3 (6 + \o) (2 + 2 D + \o)}, \\ 
\lambda_{\mathrm{c}2}^3 = \frac{3 (D - E) (1 + w_\mathrm{om})}{2 D}, \  
\lambda_{\mathrm{c}2}^4 = \frac{3 (1 + w_\mathrm{om})}{D}, $ \\
where $F = D^4 (6 + \o) {(2 + 2 D + \o)}^2 (32 D^3 (-1 + 3 w_\mathrm{om}) + 
         D^2 (34 - 42 w_\mathrm{om} (10 + 3 w_\mathrm{om}) + 
            3 \o (-1 + w_\mathrm{om}) (7 + 9 w_\mathrm{om})) - 
         3 {(1 + w_\mathrm{om})}^2 (-54 + \o (-37 - 6 \o + 6 (2 + \o) w_\mathrm{om})) - 
         6 D (1 + w_\mathrm{om}) (-58 - 6 w_\mathrm{om} + 
            \o (-17 + w_\mathrm{om} (19 + 6 w_\mathrm{om})))).$
For fixed point 3.4.2, conditions $F_{\mathrm{c}2} > 0$ and $\o > -2 -2 D$ are required.

\subsection{Fixed points of $(x \neq 0, y=0, z=0, t \neq 0)$ type}
We obtain fixed points of types $(\mathrm{d})$ from Eqs. (21)-(24).  
\subsubsection{$ (x_{\mathrm{d}1},  y_{\mathrm{d}1},  z_{\mathrm{d}1},  t_{\mathrm{d}1\pm}) = (-\frac{\sqrt{2/3} (-2 + E)}{2 + 2 E + \o}, 0, 0, \pm \frac{ \sqrt{-(-10 - 8 E + 2 E^2 - 3 \o) (6 + \o)}}{\sqrt{3} (2 + 2 E + \o)})$}

With $\Omega_{\phi_{\mathrm{d}1}}=1$, $w_{\mathrm{m}_{\mathrm{d}1}} = \frac{2 - 18 E + 4 E^2 - 3 \o}{6 + 6 E + 3 \o}$, $w_{\phi_{\mathrm{d}1}}=\frac{2 - 18 E + 4 E^2 - 3 \o}{6 + 6 E + 3 \o}$, eigenvalues are given with the substitution $D$ $\to$ $E$ for the fixed points in the subsection 3.4.1.

\subsubsection{$ (x_{\mathrm{d}2},  y_{\mathrm{d}2},  z_{\mathrm{d}2},  t_{\mathrm{d}2\pm})=(-\sqrt{\frac{3}{2}} \frac{(1 + w_\mathrm{om})}{2 E}, 0, 0, \pm \frac{\sqrt{F_{\mathrm{d}2}}}{2 \sqrt{2} E \sqrt{(2 + 2 E + \o)}})$}
Here $F_{\mathrm{d}2} = E^2 (8 - 24 w_\mathrm{om}) - 
         3 (2 + \o) (-4 + \o (-1 + w_\mathrm{om})) (1 + w_\mathrm{om}) - 
         2 E (-16 + \o (-5 + 6 w_\mathrm{om} + 3 w_\mathrm{om}^2))$. With $w_{\mathrm{m}_{\mathrm{d}2}} = \frac{-1 + (-1 + E) w_\mathrm{om}}{E}$, $\Omega_{\phi_{\mathrm{d}2}}=\frac{3 (2 + \o) (1 + w_\mathrm{om}) + 2 E (7 + 3 w_\mathrm{om})}{4 E^2}$, \\  $w_{\phi_{\mathrm{d}2}}=\frac{3 (2 + \o) w_\mathrm{om} (1 + w_\mathrm{om}) + 2 E (-2 + 5 w_\mathrm{om} + 3 w_\mathrm{om}^2)}{
3 (2 + \o) (1 + w_\mathrm{om}) + 2 E (7 + 3 w_\mathrm{om})}$, eigenvalues are given with the substitution $D$ $\to$ $E$ for the fixed points in the subsection 3.4.2. \\

In the subsections 3.2-3.5, we have found fixed points with general values $E$, $D$, $w_\mathrm{om}$, and $\o$. 
Among them, for example, the $\o = -4$ case corresponds to an effective theory of string theory \cite{53b}, and the fixed point with $w_\mathrm{om}=0$, $E=-1$, $D=2$,  $\Omega_{\phi}=1$, $w_\mathrm{m}=-1$,   $w_{\phi}=-1$, and eigenvalues [$0$, $0$, $-3$, $-3$] in the subsection 3.4.1 are normally stable (Non-Hyperbolic). 
 The  $\o = -6$ case corresponds to conformally invariant models \cite{54b}, and the fixed point with $w_\mathrm{om}=0$, $E=-1$, $D=1$, $\Omega_{\phi}=1$, $w_\mathrm{m}=-1$, $w_{\phi}=-1$, and eigenvalues  [$2$, $2$, $-1$, $-1$]  in the subsection 3.4.1 is saddle. 
(The $\o > 160000$ case satisfies 
cosmological constraints and solar-system test \cite{57b}, and the fixed point with
 $w_\mathrm{om}=0$, $E=-1$, $D=1$, $\Omega_{\phi}=1$, $w_\mathrm{m}=-1$, $w_{\phi}=-1$, and all eigenvalues are negative in the subsection 3.4.1 and then this fixed point is stable.) In the next section we analyze physically meaningful, specified more fixed points.

\section{Dynamical analysis}

We consider cases with such special values 
as $E=-1$, $D=1$ (which give us $V_\mathrm{eff}(\phi) \propto \phi^{-2}$,  $V(\phi) \propto \phi^2$), and  $w_\mathrm{om}=0$  for other non-relativistic matter, i.e. ordinary matter is dust $p_{\mathrm{om}} = 0$. 
This $j=0$ case corresponds to that given in Eq. (5).
We regard both cases with the Brans-Dicke coupling constant $\o<0$ and $\o>0$. 
Specific examples with $\o=-3$ and $\o=5$ only are studied for convenience, and 
possible trajectories  from the fixed point corresponding to the (effectively) radiation-like matter dominated era to the (effectively) dark energy-dominaed era of our Universe are to be found. \\

\subsection{$\o=-3$ case}
When $\o=-3$, by using the results in the sections $3.2$-$3.5$ we obtain explicit properties of fixed points  like cosmological parameters relevent to them.\\
 Requiring the constraint, $0$ $\leq$ $\Omega_{\phi}$ $\leq$ $1$, we have written down (selected) realistic fixed points $P_{A1}$, $P_{A2,3}$, $P_{A5,6}$ in Table $1$. We represent possible paths: \\
$1$. $P_{A1}$  $\to$ $P_{A5,6}$  \\
$2$. $P_{A2,3}$ $\to$ $P_{A5,6}$  \\   
It is shown that the paths in a phase space pass well from the radiation-like matter era to dark energy-dominated era of the Universe,  as one can see in Figs. $1$ and $2$.

\begin{table}[h]
\tbl{Fixed points (for $\o=-3$, $E=-1$, $D=1$, $w_{\mathrm{om}}=0$, and $0 \leq \Omega_{\phi} \leq 1 $), their eigenvalues, and stability.}
{\begin{tabular}{@{}ccccccccc@{}} \toprule
Point & $x$ & $y$& $z$ & $t$  & Eigenvalues & Stability \\
\colrule
$P_{A1}$ & $0$  & $ \pm 1$ & $0$   & $0$  & $(3, 3, 3, 0)$  &   $unstable$     \\
$P_{A2}$ & $\frac{1}{3(\sqrt{3} - \sqrt{6})}$  & $0$  & $0$   &$0$   & $(3 , 1.8, 1.8, 1.17)$   &       $unstable$   \\
$P_{A3}$ & $-\frac{1}{3(\sqrt{3} + \sqrt{6})}$  & $0$  & $0$   &$0$   & $(6.8, -3.8, -3.8, 3)$   & $saddle$      \\ 
$P_{A5,6}$ & $\sqrt{\frac{2}{3}}$ & $0$ &$ \pm \sqrt{7}$   & $0$  & $(-7, -7, -4, -4)$   &      $stable$ \\ \botrule
\end{tabular}\label{ta1}}
\end{table}

\begin{figure}[ph]
\centerline{\includegraphics[width=2.0in]{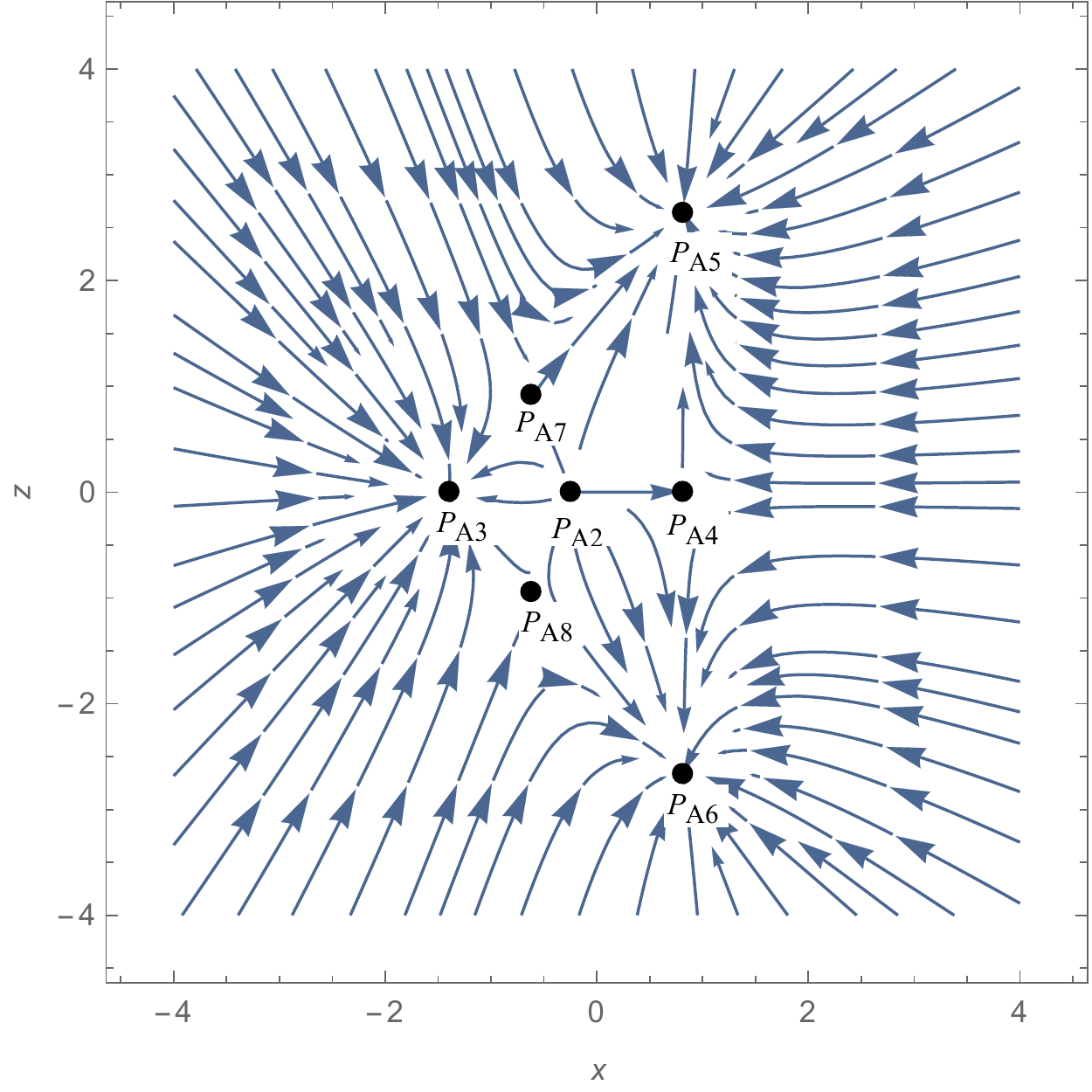}}
\vspace*{8pt}
\caption{The figure exhibits the phase space trajectories on the $xz$-plane for
the case where $j = 0$,  $\o=-3, E=-1$, and $ D=1$, among fixed points given in the 
subsection 4.1. The paths  go from $P_{A2}$ to $P_{A5, 6}$. The stable (attractor) points $P_{A5}$ and $P_{A6}$ are related to the late-time accelerating Universe, and $P_{A2}$ is an unstable point corresponding to the radiation-like matter dominated era.\protect\label{fig1}}
\end{figure}

\begin{figure}[ph]
\centerline{\includegraphics[width=2.0in]{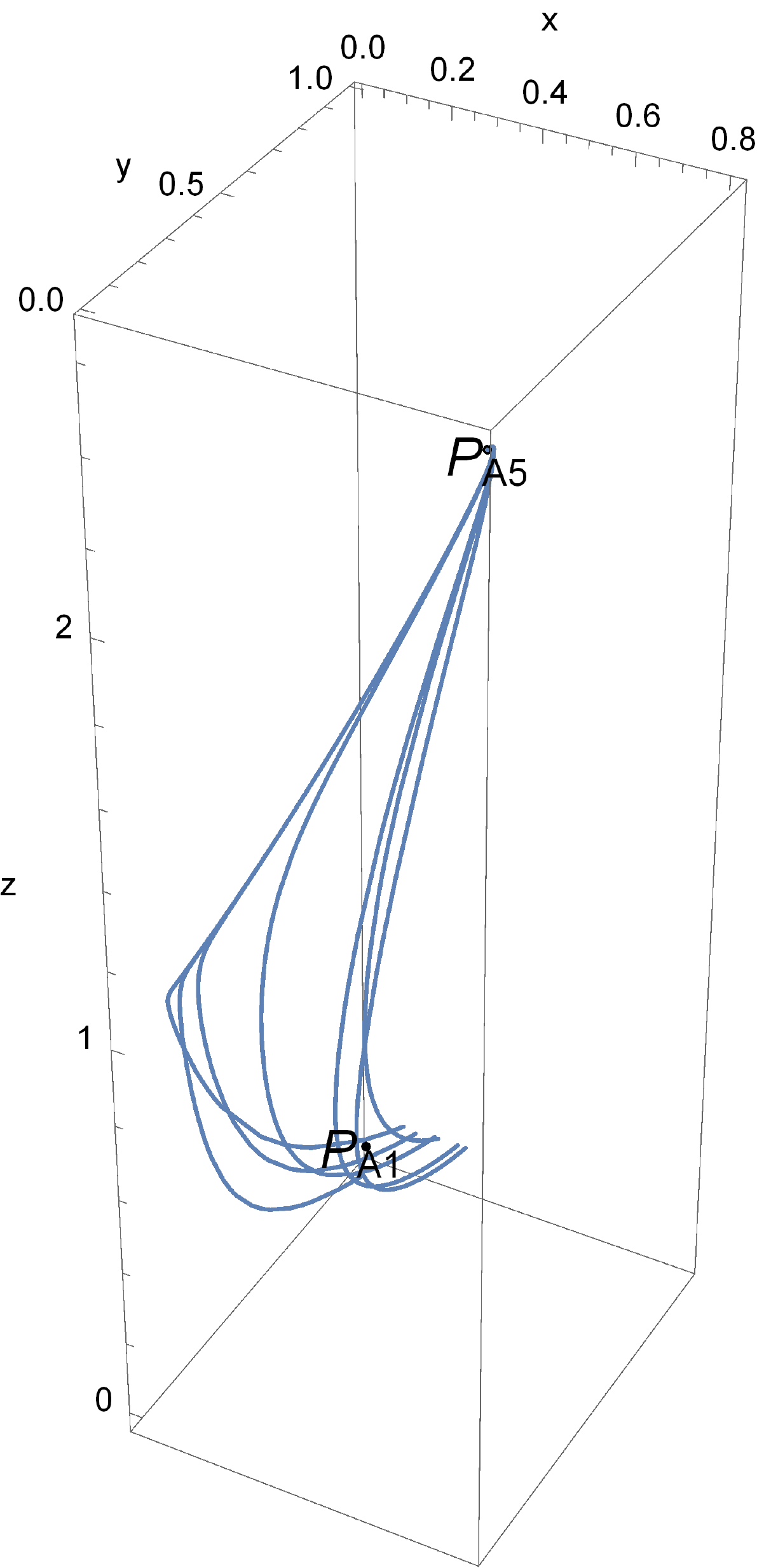}}
\vspace*{8pt}
\caption{The figure exhibits a part of phase space trajectories on the $xyz$-space for
the case where $j = 0$, $\o=-3$, $E=-1$, and $D=1$. It  shows the trajectory starting from the fixed point $P_{A1}$ is attracted toward the stable fixed point $P_{A5}$.\protect\label{fig2}}
\end{figure}

\subsection{$\o=5$ case}
When $\o=5$, we also obtain explicit properties of fixed points like cosmological parameters related to them, 
by using the results in the
 sections $3.2$-$3.5$. 
\\
Requiring the constraint, $0 \leq \Omega_{\phi} \leq 1$ again, we have realistic fixed points $P_{B1}$, $P_{B2,3}$, $P_{B5,6}$ and  $P_{B9,10}$ in Table $2$. 
We represent possible paths: \\
$3$. $P_{B1}$ $\to$ $P_{B9,10}$ $\to$ $P_{B5,6}$ \\
$4$. $P_{B2}$ $\to$ $P_{B9,10}$ $\to$ $P_{B5,6}$ \\   
We show also that these trajectories pass well from the radiation-like matter era 
to the dark energy-dominated era (
as $\Lambda CDM$ cosmological model),  as one can see in Figs. $3$-$5$. 

\begin{figure}[ph]
\centerline{\includegraphics[width=2.0in]{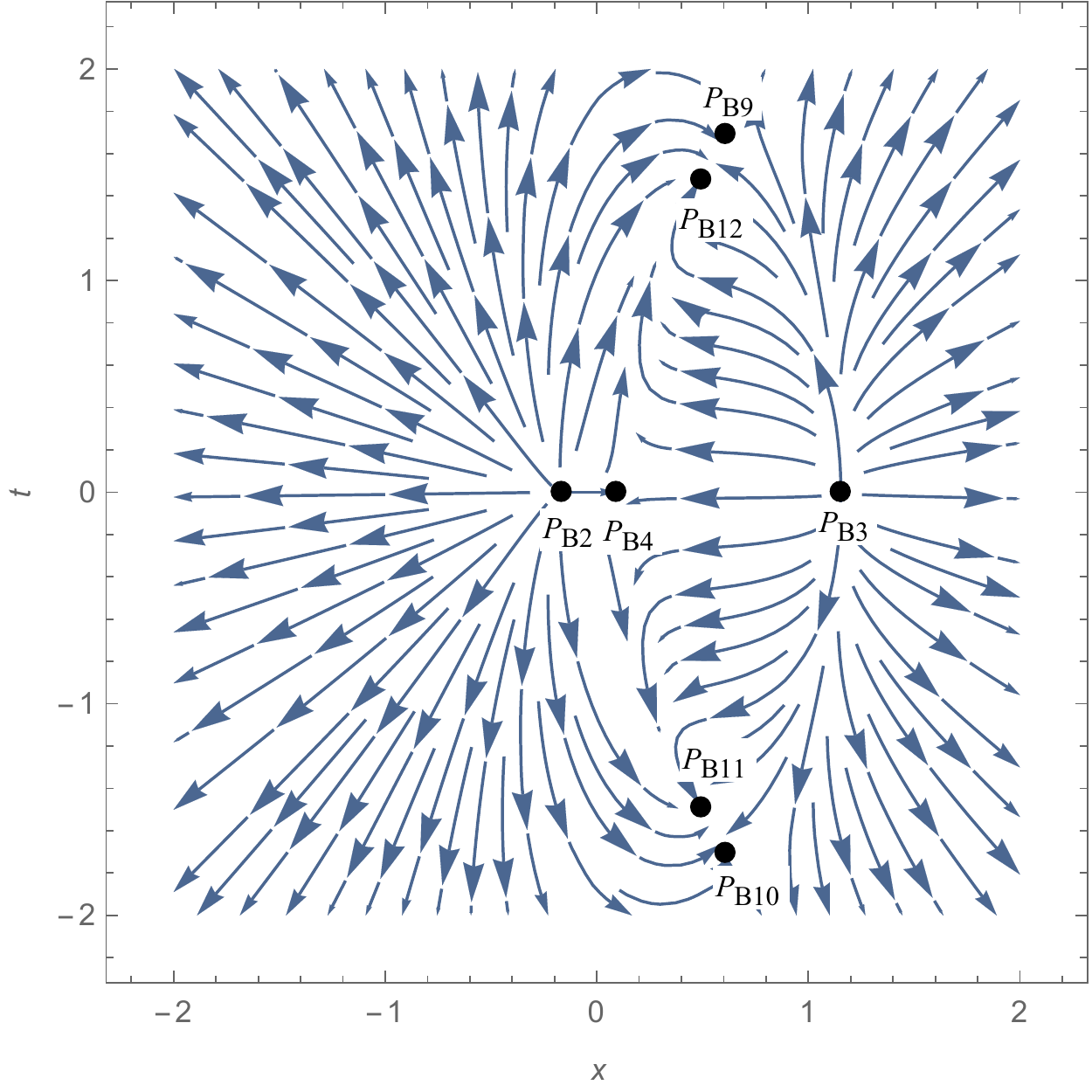}}
\vspace*{8pt}
\caption{The figure exhibits the phase space trajectories on the $xt$-space
for the case where  $j = 0$, $\o=5$, $E=-1$, and $D=1$, among fixed points given in the subsection 4.2. The paths go from $P_{B2}$ to $P_{B9,10}$. The unstable point $P_{B2}$ corresponds to the radiation-like matter dominated era.\protect\label{fig3}}
\end{figure}

 We briefly investigate the obsevational constraints in Brans-Dicke cosmology at the stable fixed points given in Tables 1 and 2. (By considering Ref. [77], we assume $\phi(\tau) = \phi_0 (\tau/\tau_0)^{\overline{\alpha}}$ and $a(\tau) =a_0 (\tau/\tau_0) ^{\overline{\beta}}$ where $\tau_0$ is current cosmic time, $\overline{\alpha}$ and $\overline{\beta}$ are constant, and we have obtained $\tau_0$ $=$ $\sqrt{\overline{\beta} (\o+6)/(3\Omega_{\mathrm{m}})}$ $H_0^{-1}$ with relation to $\overline{\alpha} = (2-2\overline{\beta})/3$, which is dependent on the Brans-Dicke parameter.) The variability\footnote{$x \equiv \frac{\dot{\phi}}{\sqrt{6} H \phi} = - \frac{\dot{G}}{2 \sqrt{6} H G}$ because $\frac{\dot{G}}{G} = \frac{-2\dot{\phi}/\phi^3}{1/\phi^2} = -\frac{2\dot{\phi}}{\phi}$.} of the gravitational constant in this Brans-Dicke theory is given by $\vert \frac{\dot{G}}{G} \vert_{0}$ $\approx$ $2.9 \times 10^{-10}$/yr  for $P_{A5,6}$ and  $\vert \frac{\dot{G}}{G} \vert_{0} \approx 3.2 \times 10^{-11}$/yr for $P_{B5,6}$ in late-time Universe, respectively. These are consistent with the observational results \cite{55b}.

\begin{figure}[ph]
\centerline{\includegraphics[width=2.0in]{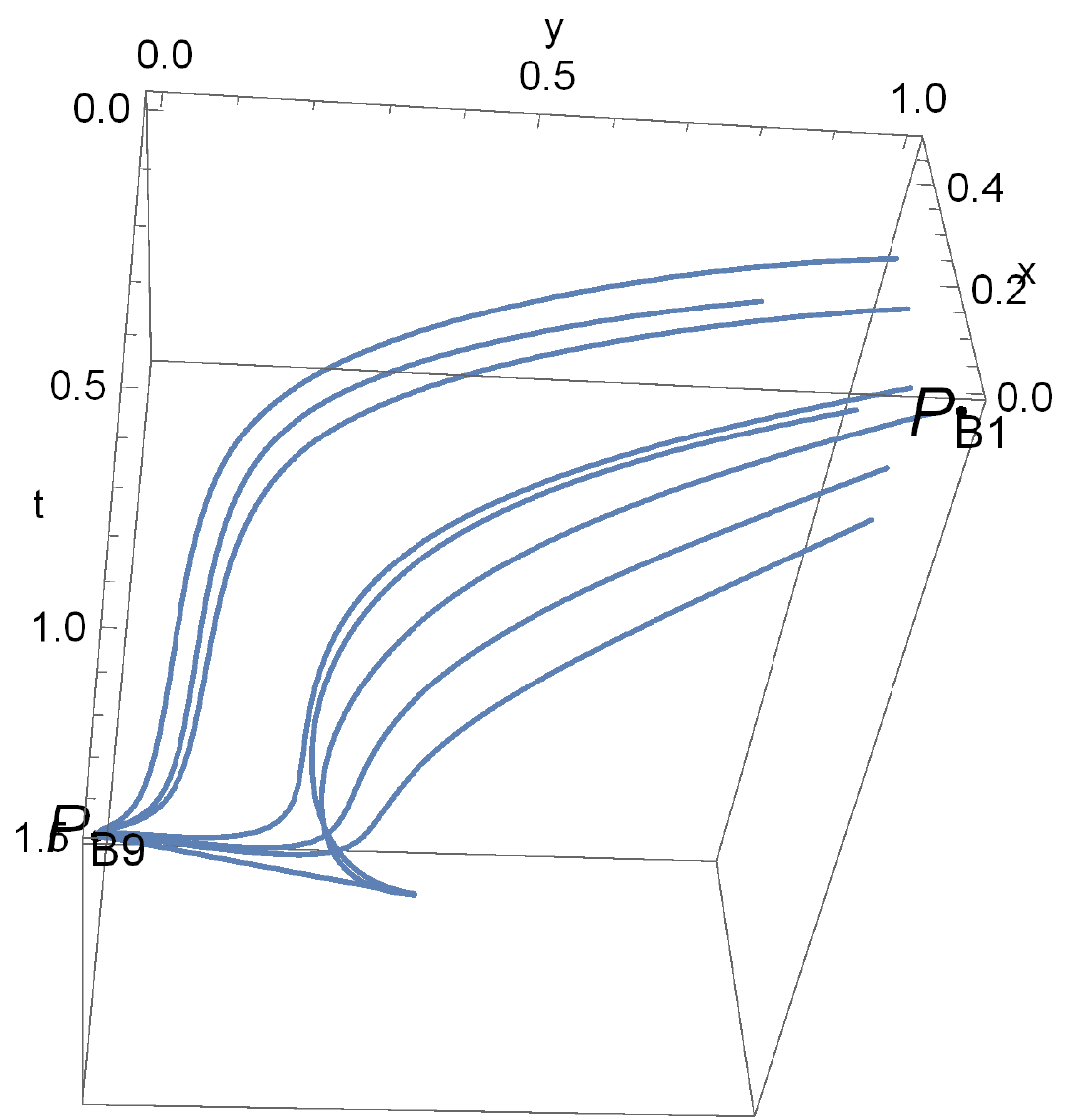}}
\vspace*{8pt}
\caption{The figure exhibits a part of phase space trajectories on the $xyt$-space for the case where $j = 0$, $\o=5$, $E=-1$, and $D=1$. The figure shows the trajectory starting from  the fixed point $P_{B1}$ is attracted toward the saddle fixed point $P_{B9}$.\protect\label{fig4}}
\end{figure}

\begin{figure}[ph]
\centerline{\includegraphics[width=2.0in]{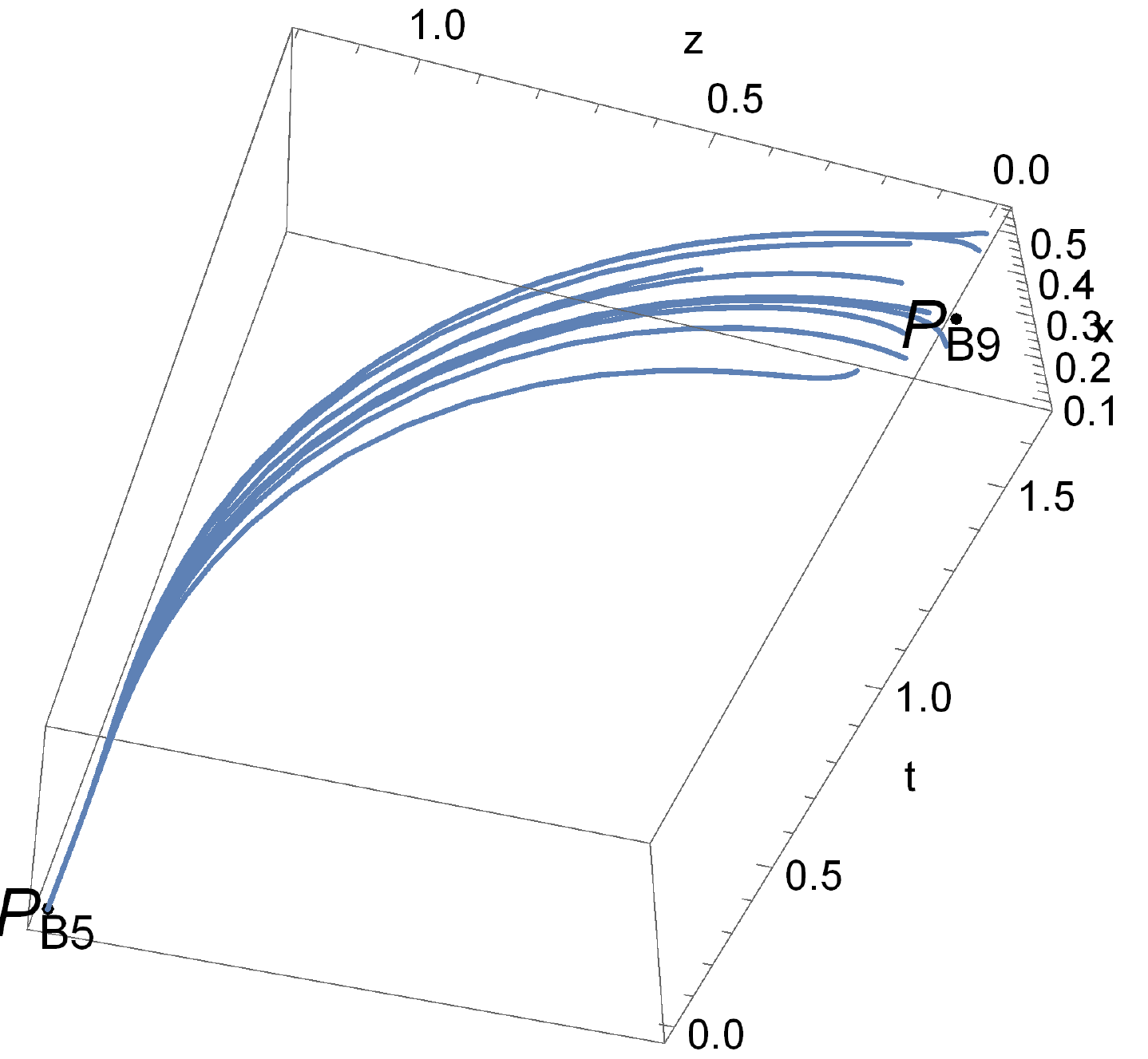}}
\vspace*{8pt}
\caption{The figure exhibits a part of phase space trajectories on
 the $xzt$-space for the case where $j = 0$, $\o=5$, $E=-1$, and $D=1$. The figure shows the  trajectory starting from  the fixed point $P_{B9}$ is attracted toward the stable fixed point $P_{B5}$.\protect\label{fig5}}
\end{figure} 

\ \ \ \ \ \ \

\begin{table}[h]
\tbl{Fixed points (for $\o=5$, $E=-1$, $D=1$,  $w_{om}=0$, and   $0 \leq \Omega_{\phi} \leq 1 $), their eigenvalues, and stability.}
{\begin{tabular}{@{}cccccccc@{}} \toprule
Point & $x$ & $y$& $z$ & $t$  & Eigenvalues & Stability \\
\colrule
$P_{B1}$  & $0$  & $\pm 1$ & $0$   & $0$  & $(3, 3, 3, 0)$  &   $unstable$      \\
$P_{B2}$  & $\frac{1}{5(\sqrt{6} - \sqrt{11})}$  & $0$  & $0$   &$0$   & $(3, 2.15, 2.15, 0.85)$   &     $unstable$       \\
$P_{B3}$  & $\frac{1}{5(\sqrt{6} + \sqrt{11})}$  & $0$  & $0$   &$0$   & $(8.6, 8.6, -5.6, 3)$   & $saddle$      \\ 
$P_{B5,6}$  & $\frac{\sqrt{\frac{2}{3}}}{9}$ & $0$ &$\pm \frac{\sqrt{\frac{341}{3}}}{9}$   & $0$  & $(-3.4, -3.4, -0.4, -0.4)$   &      $stable$  \\
$P_{B9,10}$  & $\frac{\sqrt{6}}{5}$  &  $0$ & $0$   & $\pm \sqrt{\frac{11}{5}}$  & $(-3, -2.4, 2.4, -0.6)$  &   $saddle$              \\ \botrule
\end{tabular}\label{ta2}}
\end{table}

 \ \ \ \ 
 \ \ \ \ 
\ \ \ \ \ 
 \ \ \ \ \ \ \ 
 \ \ \ \ \ \ \ 
\ \ \ \  \ \ \ 

\ \ \ \ \ \ \ 

\subsection{Dynamical system with invariant submanifolds}


In this section, we thus  investigate some invariant submanifolds to study dark energy, which make us  to exhibit more naturally the physically meaningful attractors and trajectories in 2-dimensional phase space as in Figs. 6-9.


\subsubsection{Vacuum case $\Omega_\mathrm{om}=0$}

With $\Omega_\mathrm{om}=0$, Eq. (14) becomes $1= \o x^2 - 2 \sqrt{6} x + z^2 +y^2 + t^2$. Therefore, Eqs. $(21) - (24)$ are reduced to equations of a 3 dimensional dynamical system.  For $E=-1$, $D=1$, $w_\mathrm{om}=0$, and $\o=6$, fixed points with various cosmological parameters are given below. \\

\begin{figure}[ph]
\centerline{\includegraphics[width=2.0in]{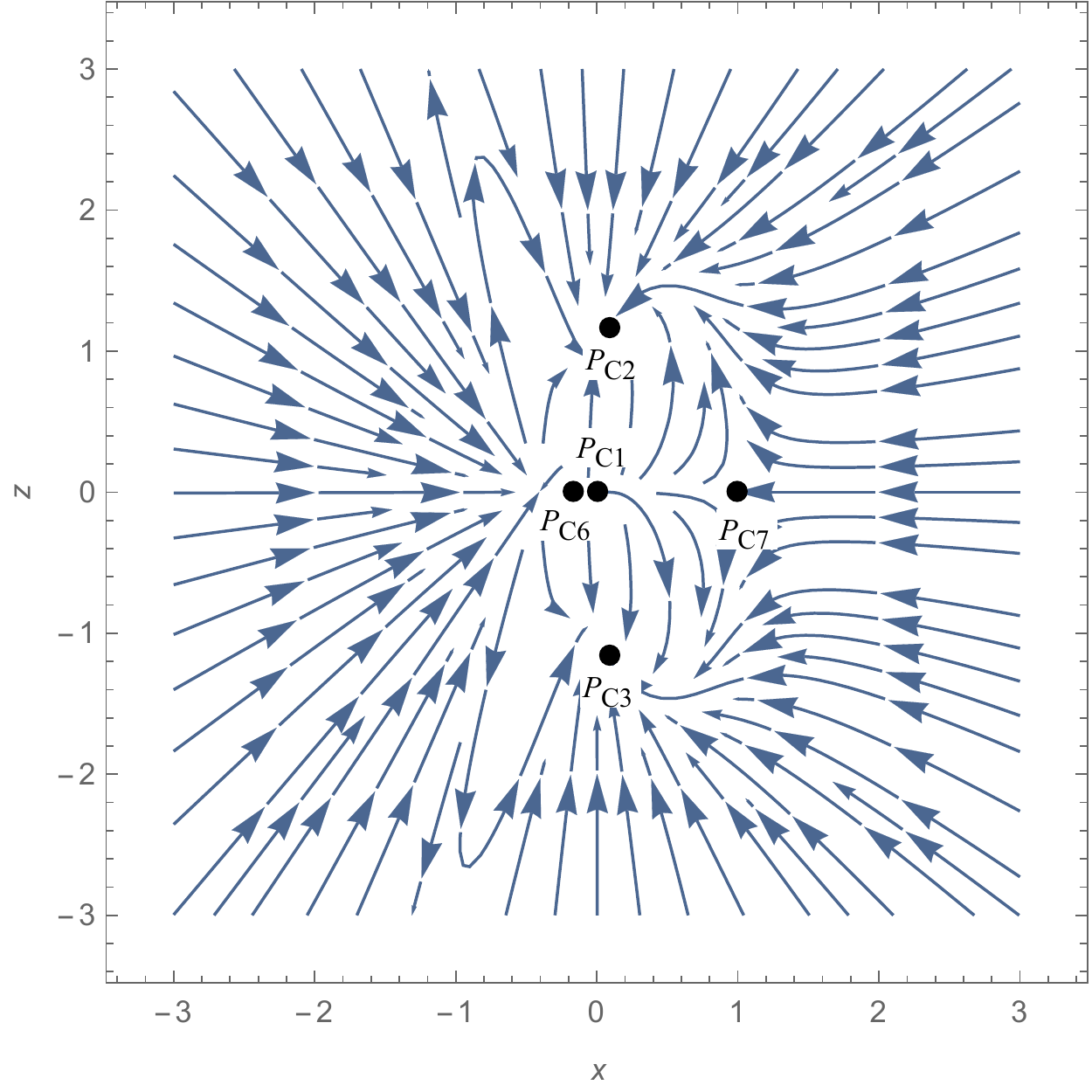}}
\vspace*{8pt}
\caption{The figure exhibits the phase space trajectory on the $xz$-plane  for the case where $w_\mathrm{om}=0$, $\o=6$, $E=-1$, and $D=1$. The figure shows the fixed point $P_{C2,3}$ is stable, corresponding to the cosmological-constant dominated era.\protect\label{fig6}}
\end{figure}

$P_{C1} = (0, 0, 0)$ : $y = 1$, $w_\mathrm{m} = w_{\phi} = 1 $. This seems to describe a stiff-matter dominated era of  Universe. The eigenvalues are $(3, 3, 0)$, which mean it is normally unstable (Non-Hyerbolic). \\ 
$P_{C2,3} = (\frac{1}{5 \sqrt{6}}, \pm \frac{\sqrt{34}}{5}, 0)$ :  $y = 0$, $w_\mathrm{m} = w_{\phi} = -1$ which seems to describe the cosmological constant dominated era of Universe. The eigenvalues are $(-\frac{4}{5}, -\frac{2}{5}, -\frac{17}{5})$, which are stable. \\
$P_{C4,5} = (\frac{1}{\sqrt{6}}, 0,  \pm \sqrt{2})$ : $y = 0$, $w_\mathrm{m}= w_{\phi} = \frac{1}{3}$ which looks like a radiation dominated era, with eigenvalues  $(-4, 2, -3)$, which are saddle. \\
$P_{C6,7} = (\frac{1}{6} (\pm 2 \sqrt{3} + \sqrt{6}), 0, 0)$ : $y = 0$, $w_\mathrm{m}= w_{\phi}= \frac{1}{18} (42 \pm 24 \sqrt{2})$, with eigenvalues   $({\frac{1}{6} (30 \pm 12 \sqrt{2}), 3, \frac{2}{3} (-6 \mp 6 \sqrt{2})})$, which are unstable(saddle). \\
This also seems a radiation dominated era.  As one can see in Fig. $6$, the paths in a phase space pass well from the radiation dominated era to dark energy dominated era of the Universe.

\subsubsection{Invariant submanifold $y = 0$ case}

When $y = 0$, we summarize various cosmological parameters of each fixed point by using the results in the sections $3.2 - 3.5$. We take $E=-1$, $D=1$, $w_\mathrm{om} = 0$ for such a reason below section. For the case $\o = 2$ fixed points with various cosmological parameters are given below. \\
 $P_{D1, 2} = (\frac{1}{3 \sqrt{6}}, \pm \frac{(2 \sqrt{\frac{11}{3}})}{3}, 0)$ :
  $\Omega_{\phi} = 1$, $w_\mathrm{m} = -1$, $w_{\phi} = -1$, eigenvalues  $(-\frac{2}{3}, -\frac{11}{3}, -\frac{11}{3})$, which are stable. \\ 
$P_{D3, 4} = (\sqrt{\frac{3}{2}}, 0, \pm 2)$ :
  $\Omega_{\phi} = 1$, $w_\mathrm{m} = 3$, $w_{\phi} = 3$, eigenvalues  $(6, -3, 3)$, which are saddle. \\
$P_{D5, 6} = (\frac{1}{2} (\pm 2 \sqrt{2} + \sqrt{6}), 0, 0)$ :   $\Omega_{\phi} = 1$,  $w_\mathrm{m} = \frac{1}{6} (30 \pm 16 \sqrt{3})$, $w_{\phi} = \frac{1}{6} (30 \pm 16 \sqrt{3})$, eigenvalues  $(\frac{1}{2} (18 \pm 8 \sqrt{3}), 3, \frac{1}{2} (18 \pm 8 \sqrt{3}))$, which are unstable. \\
$P_{D7, 8} = (- \frac{\sqrt{\frac{3}{2}}}{2}, \pm \frac{\sqrt{11}}{2}, 0)$, $\Omega_{\phi} = \frac{13}{2}$, $w_\mathrm{m} = -1$, $w_{\phi} = -\frac{2}{13}$,  eigenvalues are  $(3, - \frac{11 \sqrt{3}}{4}, \frac{11 \sqrt{3}}{4})$, which are saddle. \\
$P_{D9, 10} = (\frac{\sqrt{\frac{3}{2}}}{2}, 0, \pm \frac{\sqrt{7}}{2})$ :
 $\Omega_{\phi} = - \frac{1}{2}$,  $w_\mathrm{m} = 1$, $w_{\phi} = -2$, eigenvalues   $(3, \frac{1}{4} (-6 + 3 i \sqrt{3}), \frac{1}{4} (-6 - 3 i \sqrt{3}))$, which are saddle. \\
$P_{D11} = (\frac{1}{3 \sqrt{6}}, 0, 0)$ :
 $\Omega_{\phi} = -\frac{17}{27}$, $w_\mathrm{m} = \frac{2}{9}$, $w_{\phi} = - \frac{6}{17}$, eigenvalues  $(\frac{11}{6}, \frac{7}{6}, -\frac{11}{6})$, which are saddle. \\
Requiring the constraint, $0$ $\leq$ $\Omega_{\phi}$ $\leq$ $1$, we write down (selected) realistic fixed points $P_{D1,2}$, $P_{D3,4}$, $P_{D5,6}$ and represent possible paths. \\
5. $P_{D3,4,5,6} \to P_{D1,2}$ \\
It is shown that the paths in the phase space pass well from the radiation-like matter dominated era to dark energy  dominated era of the Universe,  as one can see in Fig. $7$.

For the case $\o = -7$ fixed points with various cosmological parameters are given below. \\ 
$P_{E1, 2} = (-\frac{\sqrt{\frac{2}{3}}}{3}, \pm \frac{\sqrt{\frac{5}{3}}}{3}, 0)$ : 
  $\Omega_{\phi} = 1$, $w_\mathrm{m} = -1$, $w_{\phi} = -1$, eigenvalues  $(\frac{4}{3}, - \frac{5}{3} , -\frac{5}{3})$, which are saddle. \\ 
$P_{E3, 4} = (-\frac{\sqrt{6}}{7}, 0, \pm \frac{1}{\sqrt{7}})$  : $\Omega_{\phi} = 1$, $w_\mathrm{m} = - \frac{15}{7}$, $w_{\phi} = - \frac{15}{7}$,  \\ eigenvalues  $(- \frac{12}{7}, -3, -\frac{33}{7})$, which are stable. \\
$P_{E5} = (- \frac{\sqrt{\frac{2}{3}}}{3}, 0, 0)$ :  $\Omega_{\phi} = \frac{22}{27}$, $w_\mathrm{m} = - \frac{4}{9}$, $w_{\phi} = - \frac{6}{11}$, eigenvalues  $(\frac{5}{6}, \frac{13}{6}, - \frac{5}{6})$, which are saddle \\
Requiring the constraint, $0$ $\leq$ $\Omega_{\phi}$ $\leq$ $1$, we  write down (selected) realistic fixed points $P_{E1,2}$, $P_{E3,4}$, $P_{E5}$ and represent possible paths. \\
6. $P_{E1,2,5} \to P_{E3,4}$ \\
It is shown that the paths in the phase space pass well from the radiation-like matter era to dark energy dominated era of the Universe,  as one can see in Fig. $8$.

\begin{figure}[ph]
\centerline{\includegraphics[width=2.0in]{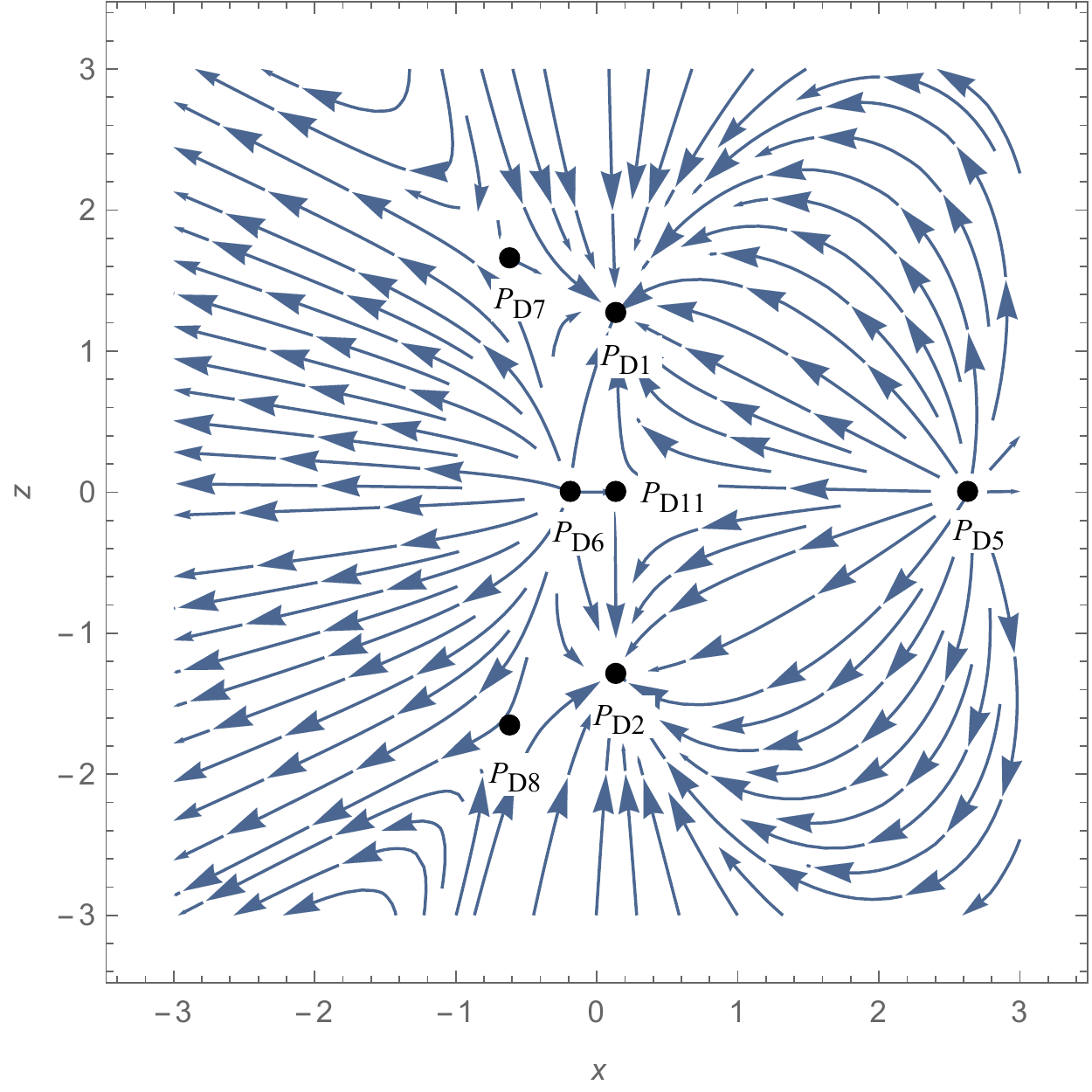}}
\vspace*{8pt}
\caption{The figure exhibits the phase space trajectory on the $xz$-plane for the case where  $w_\mathrm{om}=0$, $\o = 2$, $E=-1$, and $D=1$. The figure shows  the fixed point $P_{D1, 2}$ is stable, corresponding to the cosmological-constant dominated era. \protect\label{fig7}}
\end{figure}

\begin{figure}[ph]
\centerline{\includegraphics[width=2.0in]{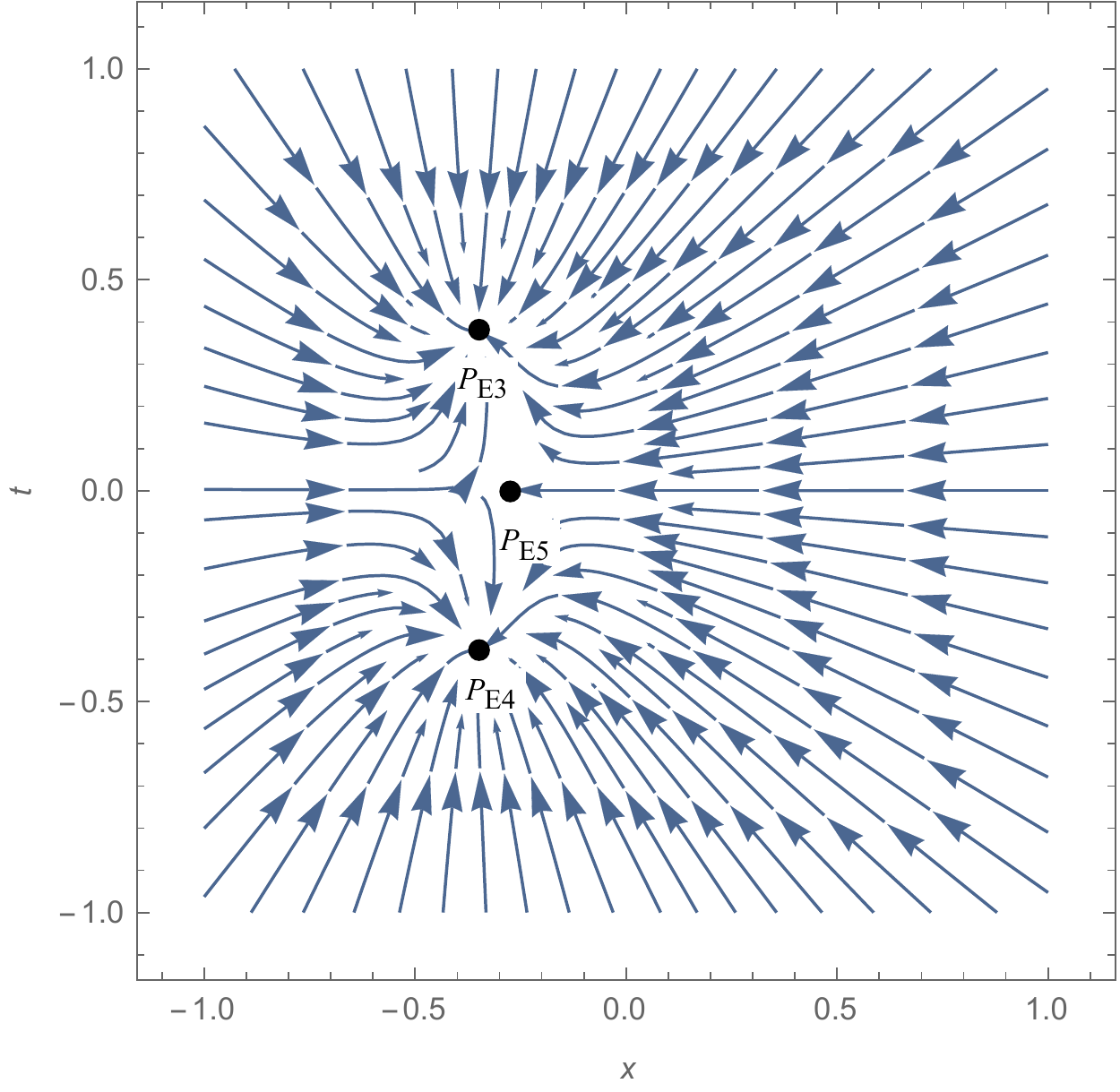}}
\vspace*{8pt}
\caption{The figure exhibits the phase space trajectory on the $xt$-plane for the case where $w_\mathrm{om}=0$, $\o = -7$, $E=-1$, and $D=1$. The figure shows  the fixed point $P_{E3, 4}$ is stable,  corresponding to a phantom dark energy.\protect\label{fig8}}
\end{figure}

\subsubsection{Invariant submanifold $x=0$ and $z=0$ case}
When $x=0$ and $z=0$, we summarize various cosmological parameters of each fixed point by using the results in the sections 3.2 - 3.5 which are reduced to 2 dimensional system. We summarize various cosmological parameters of each fixed point below \\
$P_{F1} = (0, 0)$ : $\Omega_{\phi} = 0$.  This is the ordinary matter dominated case with $w_\mathrm{m} = w_\mathrm{om}$, $w_{\phi} =$ indeterminate. Eigenvalues are $(\frac{3 (w_\mathrm{om}-1)}{2},  \frac{3 (w_\mathrm{om}+1)}{2})$ and the condition for stability is $w_\mathrm{om} < -1$. \\
$P_{F2, 3} = (\pm 1, 0 )$ :  $\Omega_{\phi} = 1$, $w_\mathrm{m} = 1$, $w_{\phi} = 1$, eigenvalues  $(3,3-3 w_\mathrm{om})$, which are unstable(saddle). \\ 
$P_{F4, 5} = (0, \pm 1)$ : $\Omega_{\phi} = 1$, $w_\mathrm{m} = -1$, $w_{\phi} = -1$, eigenvalues  $(-3,-3 (w_\mathrm{om}+1))$. The condition for stability that all eigenvalues are negative are $w_\mathrm{om}>-1$. \\
Requiring the constraint, $0$ $\leq$ $\Omega_{\phi}$ $\leq$ $1$, we  write (selected) realistic fixed points $P_{F1}$, $P_{F2, 3}$, $P_{F4,5}$ and represent possible paths. \\
7. $P_{F1,2,3} \to P_{F4,5}$ \\
It is shown that the paths in the phase space pass well from the radiation-like matter era to dark energy dominated era of the Universe,  as one can see in Fig. $9$.

\begin{figure}[ph]
\centerline{\includegraphics[width=2.0in]{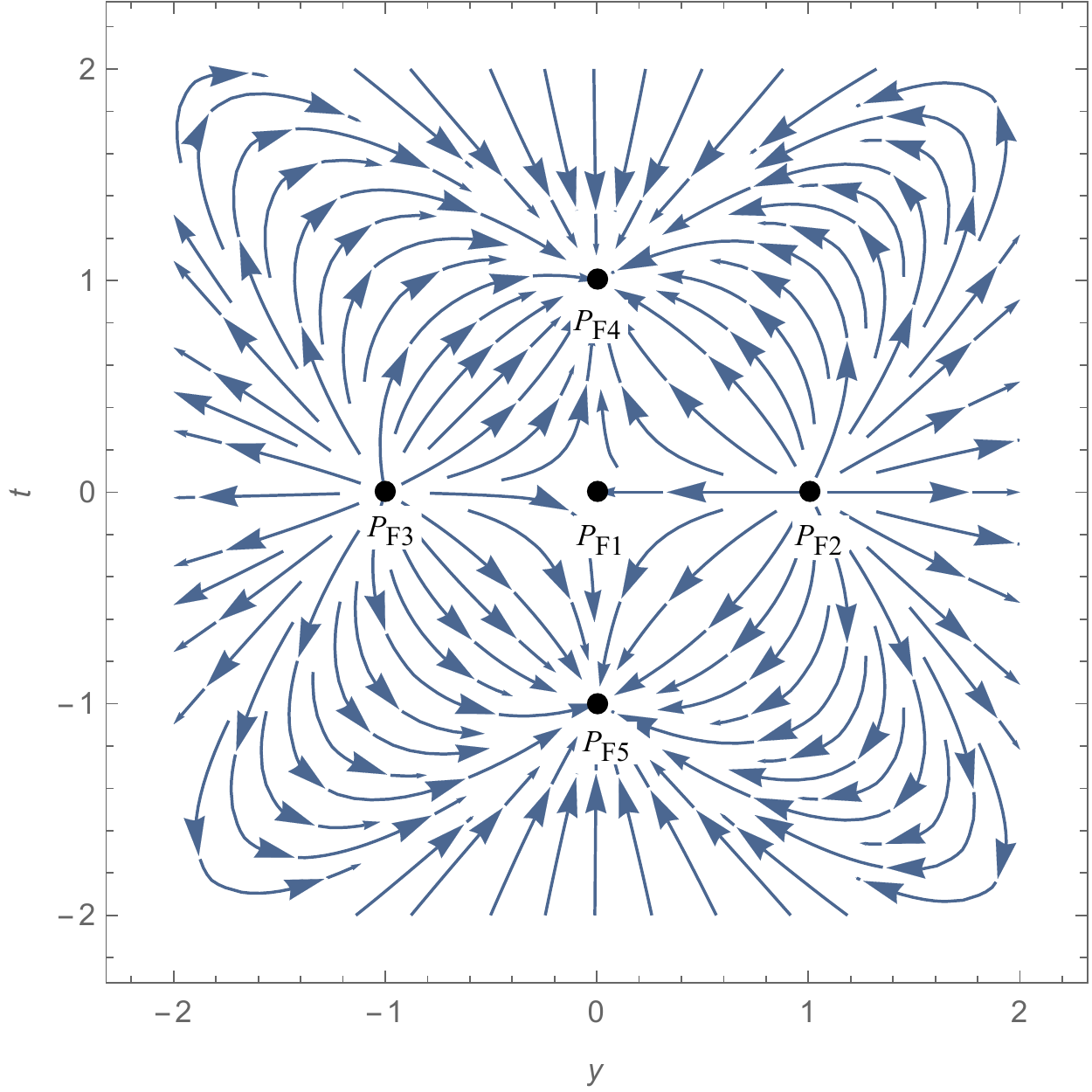}}
\vspace*{8pt}
\caption{The figure exhibits the phase space trajectory on the $yt$-plane  for the case where  $\o$ is arbitrary, $w_\mathrm{om} =0$, $E=-1$, and $D=1$. The figure shows the the fixed point $P_{F4,5}$ is stable,  mimicking the cosmological constant. \protect\label{fig9}}
\end{figure}

\section{de Sitter Universe }
In this section, we study the stability of de Sitter Universe, in which the cosmological scale factor has an exponential form.  When $A(\neq 0)$ is a constant and $B=0$, the fixed points with $x_0=0$ and $y_0=0$ satisfy Eqs. (21)-(26), which are now reduced to
\begin{eqnarray}
&& x' = A x  - \sqrt{6} x^2  , \\ 
&& y' = A y -3 \sqrt{6} x y  , \\
&& z' = \sqrt{6} D  x z - \sqrt{6} x z , \\
&& t' = \sqrt{6} E  x t - \sqrt{6} x t , \\
&& D'= 2\sqrt{6}  D^2 x [\frac{1}{2D} + \Gamma -1], \\
&& E'= 2 \sqrt{6}  E^2 x [\frac{1}{2E} +\Theta -1].
\end{eqnarray}

The equation of state regarding to the Brans-Dicke field $w_{\phi}=-1$, the  equation of state for the total energy density and pressure $w_\mathrm{m}= -z^2 - t^2$ with no other matter $w_\mathrm{om}=0$, and the density ratio regarding to the Brans-Dicke field  $\Omega_{\phi} = z^2 + t^2$.  Note that $B$ can be written as
\begin{equation}
  B \equiv - \frac{\dot{H}}{H} = \frac{3}{2}(1+ w_\mathrm{m}) 
\end{equation}
from Eqs. (7) and (8).
The fact that $B = 0$ in de Sitter spacetime is consistent with $w_\mathrm{m} = -1$ and  $z^2 + t^2 =1$. 
 From Eqs. (29)-(32) we can obtain eigenvalues of this fixed point, $(A,A, 0,0)$. When we assume that (with $B=0$, $x=0$, and $y=0$)  
\begin{equation}
a(\tau)= a_0 e^{H_0 \tau} \ , \ \phi(\tau) = \phi_0 \tau^f 
\end{equation}
where $H_0$ and $f$ are  constants, $f=0$ because $x \equiv \dot{\phi} / {\sqrt{6}H \phi}$ ($\propto$ $f/\tau$)$=0$ and
$y\equiv \sqrt{c} \dot{\phi}/{H \phi^3}$ ($\propto$ $f \tau^{-2f-1}$)$=0$. Since $A\equiv \ddot{\phi}/{H \dot{\phi}} = (f-1)/{H \tau} $, $A$ becomes  $-1$ in the late-time when $H_0 \simeq \tau_0^{-1}$ is used with the age of the Universe $\tau_0$. Thus the eigenvalues are normally stable (Non-Hyperbolic). Therfore, de Sitter Universe has solutions of the late-time attractor for the dark energy, which is dependent on the potentials. 

\section{Conclusions}

We have studied cosmology in a Brans-Dicke gravity theory with the inverse power-law potential derived from the low-energy effective theory formalism \cite{49b,50b,51b}, by using the dynamical system method.
Analyzing the evolution of our Universe as a dynamical system, we have got fixed points with  various values of the  cosmological parameters, $E, D, \o$, and $w_\mathrm{om}$, 
in the sections 3 and 4. 
  We have investigated the stability around the fixed points 
  when  $\o>0$  and $\o<0$ also. 
  In the special $\o=-3$  and $\o=5$ cases, we have described in phase spaces the evolution of the whole Universe from (unstable fixed point)  the radiation-like matter 
 to the (stable) dark energy dominated era.
  In addition, we have shown in Footnote $\mathrm{c}$ of the section 4 that a theoretical constraint for the variability $x$ of the gravitational coupling constant in our Brans-Dicke  theory is in good agreement with the experimental results \cite{55b}. 
  \\
In the section 4.3, we have studied dynamical system with an invariant submanifold such as vacuum case $\Omega_\mathrm{om}=0$. Moreover, we have analyzed specific dynamical systems such as $x=0$, $y=0$, and $z=0$ case. We have got  stable fixed points $P_{C2,3}, P_{D1,2}$ composed of only the kinetic and potential terms for Brans-Dicke field. The stable point $P_{E3,4}$ seems to correspond to a phantom dark energy composed of the kinetic term for the Brans-Dicke field and the effective potential  for the scalar field. The $x=0$ and $z=0$ case with stable point $P_{F4,5}$ can be thought as quintessence-like model, composed of effective kinetic and potential terms for the  scalar field. 
  \\
  In the section 5, for the specific, cosmic solution (with an arbitrary $\o$-value) which corresponds to de Sitter Universe we have demonstrated that it is the stable fixed point corresponding to the late-time Universe.
 \\
  In summary, we have shown that our cosmological model in a Brans-Dicke theory with inverse power-law potentials derived from the low-energy effective theory formalism 
  can describe well the late-time  Universe dominated by dark energy
 as a stable fixed point, which is 
evolved 
from the radiation-like matter dominated era (unstable fixed point).
 It would be interesting to perform sophisticated analyses with more general cases including $j\neq 0$ inverse power-law and exponential effective potentials 
 as well as a more detailed comparison to recent cosmological observations.

\section{Acknowledgements}
The authors would like to thank an anonymous reviewers for valuable comments.
 It was supported by the Basic Science Research Program through the National Research
Foundation of Korea (NRF) funded by Ministry of Education, Science and Technology 
(NRF-2017R1D1A1B06032249).


\begin{thebibliography}{0}    

\bibitem{1b} A. G. Riess et al., Astron. J. {\bf 116}, 1009 (1998)
\bibitem{2b} S. Perlmutter et al., Astrophys. J. {\bf 517},  565 (1999)
\bibitem{3b} G. F. Smoot et al., Astrophys. J. {\bf 396}, L1 (1992)
\bibitem{4b} E. Komatsu et al., Astrophys. J. Suppl. Ser {\bf 192}, 18 (2011)
\bibitem{5b} P. D. Mannheim, Astrophys. J. {\bf 479}, 659 (1997)
\bibitem{6b} T. H. Lee, B. J. Lee, Phys. Rev. D {\bf 69}, 127502  (2004) 
\bibitem{7b} P. A. R. Ade et al., Astron. Astrophys. {\bf 594}, A13 (2016)
\bibitem{8b} J. Lee, J. M. Overduin, T. H. Lee, P. Oh,  Phys. Rev. D {\bf 90}, 123003 (2014)
\bibitem{9b} V. Sahni, Class.Quantum Gravity {\bf 19}, 3435 (2002) 
\bibitem{10b} P. J. E. Peebles, B. Ratra, Rev. Mod. Phys. {\bf 75}, 559 (2003)
\bibitem{11b} R. R. Caldwell, R. Dave, P. J. Steinhardt, Phys. Rev. Lett. {\bf 80}, 1582 (1998)
\bibitem{12b} C. Armendariz-Picon, V. Mukhanov, P. J. Steinhardt,  Phys. Rev. Lett. {\bf 85}, 4438 (2000)
\bibitem{13b} A. Sen, JHEP {\bf 04}, 048 (2002)  
\bibitem{14b} C. Brans, R. H. Dicke, Phys. Rev {\bf 124}, 925 (1961) 
\bibitem{15b} E. J. Copeland, M. Sami, S. Tsujikawa, Int. J. Mod. Phys. D {\bf 15}, 1753 (2006); arXiv:hep-th/0603057
\bibitem{16b} S. Bahamonde, C. G. B$\ddot{o}$hmer, S. Carloni, E. J. Copeland, W. Fang, N. Tamanini, arXiv:1712.03107
\bibitem{17b} F. Bezrukov, M. Shaposhnikov, Phys. Lett. B {\bf 659}, 703 (2008)
\bibitem{18b} J. Sim, T. H. Lee, J. Korean Phys. Soc. {\bf 68}, 725 (2016)  
\bibitem{19b} J. Matsumoto, S. V. Sushkov, JCAP {\bf 11}, 047 (2015)  
\bibitem{20b} C. Germani, A. Kehagias, Phys. Rev. Lett. {\bf 105}, 011302 (2010) 
\bibitem{21b} S. Lee, J. Sim, T. H. Lee, J. Korean Phys. Soc. {\bf 64}, 611 (2014)
\bibitem{22b} M. A. Skugoreva, S. V. Sushkov, A. V. Toporensky, Phys. Rev. D {\bf 88}, 083539 (2013) 
\bibitem{23b} Y. Huang, Q. Gao, Y. Gong, Eur. Phys. J. C {\bf 75}, 143 (2015)
\bibitem{24b} L. Amendola., Phys. Rev. D {\bf 62}, 043511 (2000)
\bibitem{25b} A. A. Sen, S. Sethi, Phys. Lett. B {\bf 532}, 159 (2002) 
\bibitem{26b} Z. K. Guo, Y. S. Piao, Y. Z. Zhang, Phys. Lett. B {\bf 568}, 1 (2003) 
\bibitem{27b} A. de la Macorra, G. Piccinelli, Phys. Rev. D {\bf 61}, 123503 (2000)
\bibitem{28b} S. C. C. Ng, N. J. Nunes, F. Rosati, Phys. Rev. D {\bf 64}, 083510 (2001)
\bibitem{29b} W. Zimdahl, D. Pav$\acute{o}$n, L. P. Chimento, Phys. Lett. B {\bf 521}, 133 (2001)
\bibitem{30b} E. J. Copeland, A. R. Liddle, D. Wands, Phys. Rev. D {\bf 57}, 4686 (1998)
\bibitem{31b} Y. Gong, Phys. Lett. B {\bf 731}, 342 (2014)
\bibitem{32b} T. Barreiro, E. J. Copeland, N. J. Nunes, Phys. Rev. D {\bf 61}, 127301 (2000)
\bibitem{33b} D. J. Holden, D. Wands, Phys. Rev. D {\bf 61}, 043506 (2000)
\bibitem{34b} O. Hrycyna, M. Szydłowski, M. Kamionka, Phys. Rev. D {\bf 90}, 124040 (2014)
\bibitem{35b} O. Hrycyna, M. Szydłowski, Phys. Rev. D {\bf 88}, 064018 (2013)
\bibitem{36b} G. Papagiannopoulos, J. D. Barrow, S. Basilakos, A. Giacomini, A. Paliathanasis, Phys. Rev. D {\bf 95}, 024021 (2017)
\bibitem{37b} X. M. Liu, Z. X. Zhai, K. Xiao, W. B. Liu,  Eur. Phys. J. C {\bf 72}, 2057 (2012)
\bibitem{38b} A. Cid, G. Leon, Y. Leyva, JCAP {\bf 02}, 027 (2016)
\bibitem{39b} N. Roy, N. Banerjee, Phys. Rev. D {\bf 95}, 064048 (2017)
\bibitem{40b} L. M. Reyes, S. E. P. Bergliaffa, Eur. Phys. J. C {\bf 78}, 17 (2018) 
\bibitem{41b} O. Hrycyna, M. Szydłowski, JCAP {\bf 12}, 016 (2013)
\bibitem{42b} A. Paliathanasis, M. Tsamparlis, S. Basilakos, J. D. Barrow, Phys. Rev. D {\bf 91}, 123535 (2015)
\bibitem{43b} O. Hrycyna, M. Szydłowski, JCAP {\bf 11}, 013 (2015)
\bibitem{44b} L. N. Granda, D. F. Jimenez, Eur. Phys. J. C {\bf 77}, 679 (2017) 
\bibitem{45b} F. F. Bernardi, Ricardo C. G. Landim, Eur. Phys. J. C {\bf 77}, 290 (2017)
\bibitem{46b} M. Shahalam, S. D. Pathak, S. Li, R. Myrzakulov, A. Wang, Eur. Phys. J. C {\bf 77}, 686 (2017)
\bibitem{47b} V. Gupta, R. Kabir, A. Mukherjee, D. Lohiya, Int. J. Mod. Phys. D {\bf 24}, 1550068 (2015); arXiv:1501.01604
\bibitem{48b} N. Frusciante, M. Raveri, A. Silvestri, JCAP {\bf 02}, 026 (2014)
\bibitem{key-1}G. Leon, J. Saavedra, E. N. Saridakis, Class. Quantum Gravity \textbf{30}, 135001 (2013)
\bibitem{key-2}D. S. Odintsov, K. V. Oikonomou, Phys. Rev. D \textbf{98}, 024013 (2018)
\bibitem{key-3}X. Chen, E. N. Saridakis, G. Leon, JCAP \textbf{07}, 005 (2012)
\bibitem{key-4}D. S. Odintsovm, K. V. Oikonomou, Phys. Rev. D \textbf{97},
124042 (2018)
\bibitem{key-5} Ricardo C. G. Landim, Int. J. Mod. Phys. D \textbf{24},1550085 (2015) 
\bibitem{key-6}J. Dutta, W. Khyllep, E. N. Saridakis, N. Tamanini, S. Vagnozzi, JCAP \textbf{02}, 041 (2018)
\bibitem{key-7}H. Zonunmawia, W. Khyllep, N. Roy, J. Dutta, N. Tamanini, Phys. Rev. D \textbf{96}, 083527 (2017)
\bibitem{key-8}G. Kofinas, G. Leon, E. N. Saridakis, Class. Quantum Gravity \textbf{31}, 175011 (2014)
\bibitem{key-9}M. Skugoreva, E. N. Saridakis, A. Toporensky, Phys. Rev. D \textbf{91}, 044023 (2015)
\bibitem{key-10} Carlos R. Fadragas, G. Leon, E. N. Saridakis, Class. Quantum Gravity \textbf{31}, 075018 (2014)
\bibitem{key-11}V. K. Oikonomou, Int. J. Mod. Phys. D. \textbf{27}, 1850059 (2018) 
\bibitem{key-12}S. D. Odintsov, V. K. Oikonomou, Phys. Rev. D \textbf{96}, 104049 (2017)
\bibitem{key-13} Ricardo C. G. Landim, Eur. Phys. J. C \textbf{76}, 31 (2016)
\bibitem{key-14}J. Dutta, W. Khyllep, N. Tamanini, JCAP \textbf{01}, 038 (2018)
\bibitem{key-15}G. Leon, E. N. Saridakis, JCAP \textbf{03}, 025 (2013)
\bibitem{key-16}G. Leon, E. N. Saridakis, JCAP \textbf{11}, 006 (2009)
\bibitem{key-17}G. Leon, E. N. Saridakis, Class. Quantum Gravity \textbf{28}, 065008 (2011)
\bibitem{key-18} Ricardo C. G. Landim, Eur. Phys. J. C, \textbf{76}, 480 (2016)
\bibitem{key-19}G. Leon, E. N. Saridakis, JCAP \textbf{04}, 031 (2015)
\bibitem{key-20}X. Chen, Y. Gong, E. N. Saridakis, JCAP \textbf{04}, 001 (2009)
\bibitem{49b} B. Damdinsuren, J. Sim, T. H. Lee, Class. Quantum Gravity {\bf 34}, 175012 (2017)
\bibitem{50b} C. Lee, T. H. Lee, H. Min, Phys. Rev. D {\bf 39}, 1681 (1989)
\bibitem{51b} C. Lee, T. H. Lee, H. Min, Phys. Rev. D {\bf 39}, 1701 (1989)
\bibitem{52b} A. Avilez, C. Skordis, Phys. Rev. Lett {\bf 113}, 011101 (2014)
\bibitem{53b} V. Faraoni, \textit{Cosmology in Scalar-Tensor Gravity}, Fundamental Theories of Physics (Kluwer Academic Publishers, Dordrecht, Boston, 2004), Vol. 139
\bibitem{54b} S. Deser, Ann. Phys. NY {\bf 59}, 248 (1970) 
\bibitem{55b} J. P. Uzan, Living Rev. Rel {\bf 14}, 2 (2011)
\bibitem{57b} Bertotti, B., Iess, L., Tortora, P., Nature 425, 374 (2003)
\bibitem{58b} O. Bertolami, P.J. Martins, Phys. Rev. D {\bf 61}, 064007 (2000) 


\end{thebibliography}
\end{document}